\documentclass[a4paper]{article}
\usepackage{indentfirst}
\usepackage{hyperref}
\usepackage{latexsym,bm,amsmath,amssymb}
\usepackage{graphicx}
\usepackage{epsfig}
\usepackage{epstopdf}
\usepackage{subfigure}
\usepackage{color}
\usepackage[top=1in, bottom=1in, left=1.2in, right=1.2in]{geometry}
\usepackage{appendix}
\usepackage{ulem}

\newcommand{\zc}{Z_c(3900)}

\begin{document}

\title{ $Z_c(3900)$ as a $\bar D D^*$ molecule from the pole counting rule}

\author{ Qin-Rong~Gong$^{1}$\thanks{1201110067@pku.edu.cn}, Zhi-Hui ~Guo$^{2,3}$\thanks{zhguo@hebtu.edu.cn}, Ce ~Meng$^{4}$\thanks{mengce75@pku.edu.cn}, \\
Guang-Yi~Tang$^{5}$\thanks{tangguangyi@foxmail.com}, Yu-Fei Wang$^{1}$\thanks{wyf19910927@126.com}, Han-Qing~Zheng$^{1,6}$\thanks{zhenghq@pku.edu.cn}
\vspace*{0.3cm} \\
$^{1}${\it Department of Physics and State Key Laboratory of Nuclear Physics and Technology,} \\
{\it Peking University,  Beijing 100871, China}\\
$^{2}${\it  Department of Physics, Hebei Normal University, Shijiazhuang 050024, China} \\
$^{3}${\it  Helmholtz-Institut f\"ur Strahlen- und Kernphysik and} \\
  {\it Bethe Center for Theoretical Physics,  Universit\"at Bonn, D-53115 Bonn, Germany} \\
$^{4}${\it  Department of Physics, Peking University, Beijing 100871,  China} \\
$^{5}${\it  Institute of High Energy Physics, Beijing 100049,  China} \\
$^{6}${\it  Collaborative Innovation Center of Quantum Matter, Beijing 100871, China}
}

\maketitle

\begin{abstract}
A comprehensive study on the nature of the $Z_c(3900)$ resonant structure is carried out in this work. By constructing the pertinent effective Lagrangians and considering the important final-state-interaction effects, we first give a unified description to all the relevant experimental data available, including the $J/\psi\pi$ and $\pi\pi$ invariant mass distributions from the $e^+e^-\to J/\psi\pi\pi$ process, the $h_c\pi$ distribution from $e^+e^-\to h_c\pi\pi$ and also the $D\bar D^{*}$ spectrum in the $e^+e^-\to D\bar D^{*}\pi$ process. After fitting the unknown parameters to the previous data, we search the pole in the complex energy plane and find only one pole in the nearby energy region in different Riemann sheets. Therefore we conclude that $Z_c(3900)$ is of $D\bar D^*$ molecular nature, according to the pole counting rule method~[Nucl.~Phys.~A543, 632 (1992); Phys.~Rev.~D
35,~1633 (1987)]. We emphasize that the conclusion based upon the pole counting method is not trivial, since both the $D\bar D^{*}$ contact interactions and the explicit $Z_c$ exchanges are introduced in our analyses and they lead to the same conclusion.
\end{abstract} 
\newpage

\section{Introduction}\label{intro}

The discovery of $X(3872)$ has opened a new era in hadron physics~\cite{Choi:2003ue}.
Since then more than two dozen of the so-called exotic $XYZ$ particles with hidden heavy-quark flavors have been observed in the past decade~\cite{Agashe:2014kda}, see Refs.~\cite{Olsen:2014qna,Chen:2016qju} for recent experimental and theoretical reviews respectively. A common and interesting feature shared by these exotic states is that many of them lie close to the underlying thresholds composed of the open heavy-flavor states. Then one important question is whether these newly observed $XYZ$ peaks from experimental analyses correspond to the genuine or elementary resonance states, or the nearby threshold effects, or even the mixture of the previous two mechanisms.
Many different methods have been proposed to discern the inner structures of the hadronic states, including the compositeness coefficient analyses~\cite{Weinberg:1962hj,Baru:2003qq,Hyodo,Aceti,Agadjanov:2014ana,Guo:2015daa},
the QCD sum rule study~\cite{Chen:2015ata,Zhang:2013aoa,Wang:2013vex,Cui:2013yva}, the $N_C$ trajectories of the resonance poles~\cite{Guo:2011pa,Guo:2015dha,Lu:2016gev}, 
and the pole counting rule~\cite{Morgan:1992ge,Zhang:2009bv,Meng:2014ota,Dai:2012pb}. In the present work, we shall apply the latter approach to study the $Z_c(3900)$, which was first observed by BESIII Collaboration~\cite{Ablikim:2013mio}.

The charged charmoniumlike state $Z_c(3900)$, with mass $3899.0\pm 3.6\pm 4.9$~MeV and width {$46\pm 10\pm 20
\mbox{MeV}$}, has been observed in the $J/\psi\pi^{\pm}$ invariant mass spectrum in $e^+e^-\to J/\psi\pi^{+}\pi^{-}$ process by BESIII collaboration~\cite{Ablikim:2013mio} in 2013. It has been confirmed by the Belle~\cite{Liu:2013dau} and CLEO\cite{Xiao:2013iha} Collaborations in $e^+e^-\to J/\psi\pi^{+}\pi^{-}$ as well.  Later on a peak structure with mass $3883.9\pm4.5$~MeV and width $24.8\pm11.5$~MeV, was observed in the $(D\bar D^{*})^\pm$ invariant mass spectrum in the $e^{+}e^{-}\rightarrow \pi^\pm(D\bar D^*)^{\mp}$ process~\cite{Ablikim:2013xfr}. The angular
 distribution analysis on the $\pi Z_c$ system in the $\pi^\pm(D\bar D^*)^{\mp}$ channel given in Ref.~\cite{Ablikim:2013xfr} determines the quantum number of $Z_c$ to be $I(J^P)=1(1^{+})$.

Regarding the nature of $Z_c(3900)$, it is very interesting  in the theoretical aspect, ifn the sense that at the quark level it can be accommodated only by a $\bar cc$ plus a light quark-antiquark pair assignment and cannot be the conventional $\bar cc$ charmonium. If these quarks are bounded together by color confining force then  $Z_c(3900)$ would be the compact tetraquark state~\cite{Maiani2013,Dias:2013xfa,Terasaki:2013lta,Qiao:2013raa,Braaten:2013boa,Wang:2013vex,Navarra:2015rha}. On the other hand, due to the fact that the mass of $Z_c(3900)$ is very close to the $D\bar{D}^*$ threshold, it provides a natural candidate as a $D\bar D^*$ molecule~\cite{Zhang:2013aoa,Dong:2013iqa,Guo:2013sya,He:2015mja,Wang:2013cya,Wilbring:2013cha,Cui:2013yva,Guo:2013ufa,Chen:2015ata,Zhou:2015jta,Albaladejo:2015lob,Patel:2014vua}. In addition there also exist other possible explanations, such as the cusp effects suggested in Refs.~\cite{Swanson:2014tra,Chen:2013coa,Liu:2013vfa,Ikeda:2016zwx}, and the anomalous triangle singularity in  Refs.~\cite{Liu:2015taa,Szczepaniak:2015eza}. A recent lattice investigation in Ref.~\cite{Chen:2014afa} gives a small negative scattering length in the corresponding channel (repulsive interaction), hence in disfavor of a resonant picture. But one should be cautious when interpreting the lattice result with the physical measurements, since the lightest pion mass used in the previous lattice study is 300~MeV, still much larger than its physical value.
Efforts have also been made in the literature~\cite{Li:2013xia,Voloshin:2013dpa,Ke:2013gia} in an attempt to distinguish between the molecule picture and the tetraquark one, by estimating the decay width of $Z_c(3900)$. In Ref.~\cite{Lin:2013mka} the $Z_c(3900)$ in the photoproduction process $\gamma p\to Z_c(3900)^+ n$ is studied.

In this work the so-called pole counting rule, which was developed in Ref.~\cite{Morgan:1992ge}, shall be used in discriminating the nature of $\zc$. This rule provides an elegant way to distinguish the mechanisms of generating resonances around the threshold by counting the number of nearby poles of the amplitudes in the complex energy plane. Through the study of the potential scattering with and without the Castillejo-Dalitz-Dyson pole, it concludes that for the molecular type of resonance there is just one corresponding pole, while for the elementary type of resonance there is a pair of poles lying close to the threshold in the nearby unphysical Riemann sheets. Though this rule is not asserted as a mathematical theorem, it is most likely to hold in the physical regimes~\cite{Morgan:1992ge}. In order to apply this rule, a precisely determined amplitude is clearly crucial, and it also implies that the final conclusion from this approach relies heavily on the input experimental data. This is one of the reasons that we try to fit as many available data as possible to better constrain the amplitude.

In this respect, we should mention that the very recent works in Refs.~\cite{Guo:2014iya,Albaladejo:2015lob} have also analyzed the $D\bar D^*$ and $J/\psi\pi$ invariant mass distributions by emphasizing the nonperturbative nature of the $D\bar{D}^*$ interactions around the $\zc$ energy region. However we point out that in addition to the $D\bar  D^*$ resummation mechanism to include a bare $\zc$ field through the Breit-Wigner or Flatt\'e functions~\footnote{We are interested in the threshold energy region, therefore  the Flatt\'e-type function is more proper than the Breit-Wigner type with a constant width.} is another possibility to fit the data. Naively speaking, the two mechanisms correspond to rather different scenarios for $\zc$. If $\zc$ is originated from the $D\bar D^*$ resummation, it has a good chance to be a $D\bar D^*$ molecular, while if it is an elementary state with $D\bar D^*$ playing only a minor role in its composition, it can be easily described by the Flatt\'e function. It should be emphasized that a near-threshold pole location does not necessarily imply a molecular structure. A good example is the $X(3872)$ resonance~\cite{Meng:2014ota}. A Flatt\'e-type function fit with and explicit resonance exchange gives an excellent description to the data and pairs of poles  are found in nearby Riemann sheets.  Regarding $\zc$, it is a priori not known which approach gives better description of the data: the $D\bar D^*$ resummation or the explicit resonance exchange. One of the novelties in the present work is to include both of the approaches to fit the data. Moreover, instead of simply parameterizing the $D\bar D^*$ interactions with $X(4260)$, $J/\psi$ and other light-flavor mesons with constant or polynomial three-momentum terms~\cite{Guo:2014iya,Albaladejo:2015lob}, we describe all the interactions by constructing the pertinent effective Lagrangians and seriously take care of the resummation of $D\bar D^*$ by distinguishing the transverse and longitudinal parts of the vector mesons. A simultaneous description of the $h_c\pi$ and $\pi\pi$ invariant mass distributions, in addition to the $D\bar D^*$ and $J/\psi\pi$ ones, will be discussed. A comparison between different scenarios realized by switching on and off different parameters is made, the pole counting method established in Ref.~\cite{Morgan:1992ge} is then applied and we reach a solid conclusion that $Z_c(3900)$ is of a molecule nature.

 This paper is organized as follows. In Sec.~\ref{sec.theo}, we set up the theoretical framework, including the construct of the effective Lagrangians and the resummation of $D\bar D^*$ loops. The phenomenological results and discussions are presented in Sec.~\ref{sec.numeric}. A short summary and the conclusions are given in Sec.~\ref{sec.conclu}.

\section{ Theoretical framework to study $e^+e^-\to J/\psi\pi\pi, D\bar{D}^*\pi, h_c\pi\pi$ channels}\label{sec.theo}
\subsection{ Pertinent effective Lagrangians}\label{sec1}

Assuming that $Z_c(3900)$ is an $I(J^{P})=1(1^{+})$ particle, we construct the effective Lagrangian to describe the interactions between $Z_c(3900)$ and other particles. Here we use the conventional Proca vector representation to incorporate the vector and axial-vector states $X(4260)$ and $Z_c(3900)$, denoted by $X^{\mu}$ and $Z_c^{\mu}$,  respectively, while in Ref.~\cite{Dai:2012pb} the antisymmetric tensor formalism is employed.
In the molecular picture, the nonperturbative $D\bar D^*$ interaction is responsible for  the near-threshold state $\zc$.
The one-pion exchange contribution in the open heavy-flavor meson-meson scattering is still under debate, as discussed in Ref.~\cite{Guo:2016bjq}. There is evidence that the pion-exchange contribution is a subleading effect in the heavy-flavor meson-meson ($D^{(*)}D^{*}/B^{(*)}B^{*}$) scattering. For example, a perturbative nature of the pion exchange is established based on the power counting scheme of the effective field theory in Ref.~\cite{pavon.pwc.230216.1}. Especially it is obtained that the expansion energy scale for the isovector case increases by a factor of 3 compared to the isoscalar one, implying a more perturbative nature of the $D\bar{D}^*$ scattering in the $Z_c$ channel than that in the $X(3872)$ case.
This fact is further supported by explicit calculations, which show that the light $\bar qq$ meson exchanges indeed play a minor role in the $D\bar D^{*}$ scattering~\cite{Aceti:2014uea}.
Based on this argument, we will consider only the local contact $D\bar D^{*}$ interaction in this work.
The contact $D\bar D^*$ four-point interaction takes the following form under the consideration of heavy quark symmetry~\cite{mehen},
   \begin{align}
  \mathcal{L}_{DD^*DD^*}=\lambda_1\langle(D\bar{D}^{*\mu}+h.c.)^2\rangle,\label{eq1a}
  \end{align}
   where the field operators $D$ and $D^*$ are $SU(2)$ isospin doublets,
\begin{equation*}\label{ddoublet}
D=\begin{pmatrix}D^+\\D^0\end{pmatrix},D^{*}=\begin{pmatrix}D^{*+}\\D^{*0}\end{pmatrix}.
\end{equation*}
The $\bar{D}D^*$ meson loop should include the $\bar{D^0}D^{*\pm}$, $\bar{D}^{-}D^{*+}$ and $\bar{D}^0D^{*0}$ intermediate states.
To simplify the notations, $\bar{D}D^*/\bar{D}^*D$ is denoted as $\bar{D}D^*$ from now on.

We construct formally the effective Lagrangian in a relativistic framework, but we should mention that only the energy regions close to the $\bar{D}D^*$ threshold will be focused. Concerning the $\bar{D}D^*$ interactions with $J/\psi\pi$ and $h_c\pi$, the operators with the lowest number of derivatives which are invariant under $C$, $P$, chiral and isospin symmetry transformations, are given by
  \begin{align}
  &\mathcal{L}_{DD^*\psi\pi}=\lambda_2\nabla_\nu \psi_\mu\langle\bar{D}^{*\mu}u^\nu D\rangle+\lambda_{3}\psi_\mu\langle\nabla^\nu \bar{D}^{*\mu}u_\nu D\rangle,\nonumber\\
  &\qquad\qquad\,\,+\lambda_{4}\nabla_\nu \psi_\mu\langle\bar{D}^{*\nu}u^\mu D\rangle+\lambda_{5}\psi_\mu\langle\nabla^\mu \bar{D}^{*\nu}u_\nu D\rangle+h.c.,\label{eq1b}\\
  &\mathcal{L}_{DD^*h_c\pi}=(\lambda_{6}\nabla_\mu H_\nu\langle\bar{D}^*_\rho u_\sigma D\rangle+\lambda_{7}H_\mu\langle\nabla_\nu\bar{D}^*_\rho u_\sigma D\rangle)\epsilon^{\mu\nu\rho\sigma}+h.c.,\label{eq1c}
\end{align}
where $H^{\mu}$ denotes the axial-vector state $h_c$, $\psi^{\mu}$ stands for the $J/\psi$, $\nabla_\mu $ is a covariant
 derivative operator, $\epsilon^{\mu\nu\rho\sigma}$ is the fourth-order antisymmetric tensor, and $u_\mu$ corresponds to the standard chiral
building block that includes the light-flavor mesons
\begin{align*}
  &u=\exp(\frac{i\phi}{\sqrt{2}f_\pi}), \phi=\begin{pmatrix}\frac{\pi^0}{\sqrt{2}}& \pi^+\\\pi^-& -\frac{\pi^0}{\sqrt{2}}\end{pmatrix},\\
  &u_\mu = i{u^\dagger\partial_\mu u-u\partial_\mu u^\dagger}.
\end{align*}
For details of the chiral building blocks one is referred to, for example, Ref.~\cite{Bijnens:1999sh}.

On the other side, to include $Z_c(3900)$ as a bare state (i.e. the compact quark(s)/anti-quark(s) bounded together through color force), we include the following operators
\begin{align}
  &\mathcal{L}_{XZ_c\pi}=g_4\nabla_\nu X_\mu\langle Z_c^\mu u^\nu\rangle, \label{eq2a}\\
  &\mathcal{L}_{Z_c\psi\pi}=g_5\nabla_\nu \psi_\mu\langle Z_c^\mu u^\nu\rangle,\label{eq2b} \\
  &\mathcal{L}_{Z_cDD^*}=f_5\langle Z_c^\mu(D\bar{D}^*_\mu+h.c.)\rangle,\label{eq2c}\\
  &\mathcal{L}_{Z_ch_c\pi}=f_7\nabla_\nu H_\mu\langle Z_{c\rho}u_\sigma\rangle\epsilon^{\mu\nu\rho\sigma},\label{eq2d}
\end{align}
where $Z_c^\mu$ is given by a $2\times 2$ matrix
\begin{equation*}
  Z_c^\mu=\begin{pmatrix}\frac{Z_c^0}{\sqrt{2}} &Z_c^+\\Z_c^-& -\frac{Z_c^0}{\sqrt{2}}\end{pmatrix}.
\end{equation*}

The four-point interactions between $X(4260)$ and $J/\psi\pi\pi$, $h_c\pi\pi$, $\bar{D}D^*\pi$ are also taken into account
\begin{align}\label{lag}
  &\mathcal{L}_{X\psi\pi\pi}=g_1X_\mu\psi_\nu \langle u^\mu u^\nu\rangle+g_2X_\mu\psi^\mu\langle u^\nu u^\nu\rangle+g_3X_\mu\psi^\mu\langle\chi_\pm\rangle, \\
  &\mathcal{L}_{XDD^*\pi}=f_1\nabla_\nu X_\mu\langle\bar{D}^{*\mu}u^\nu D\rangle+f_2X_\mu\langle\nabla^\nu \bar{D}^{*\mu}u_\nu D\rangle \nonumber\\
  &+f_3\nabla_\nu X_\mu\langle\bar{D}^{*\nu}u^\mu D\rangle+f_4X_\mu\langle\nabla^\mu \bar{D}^{*\nu}u_\nu D\rangle+h.c.,\\
  &\mathcal{L}_{Xh_c\pi\pi}=f_6\nabla^\lambda\nabla_\rho X_\mu H_\nu\langle u_\lambda u_\sigma\rangle\epsilon^{\mu\nu\rho\sigma}\,,
\end{align}
where $\chi_{\pm}$ is another standard chiral building block,
$\chi_\pm = u^\dagger\chi u^\dagger\pm u\chi^\dagger u$~\cite{Bijnens:1999sh}.

In addition to the strong $\bar DD^*$ final state interaction (FSI), we also carefully include the $\pi\pi$ FSI in the processes $e^+e^- \to J/\psi\pi\pi$ and  $e^+e^- \to h_c\pi\pi$  , where
the $\pi\pi$ system is mainly in $s$-wave.
The strong $\pi\pi$ interactions are taken into account within the framework of unitarized chiral perturbation theory ($\chi$PT) up to next-to-leading order~\cite{Dai:2011bs}, where the $O(p^4)$ low energy constants are fixed by fitting the $\pi\pi$ scattering data. The resulting poles of the $f_0(500)$ and $f_0(980)$ are in good agreement with those from the more strict approaches~\cite{Zhou:2004ms,Caprini:2005zr}.
The details of the $\chi$PT Lagrangian and the unitarization procedure for $\pi\pi$ scattering shall not be repeated in the present work. Interested readers are referred to  Ref.~\cite{Dai:2011bs} and references therein.

Before ending this subsection, we point out that the word `` effective'' in the effective Lagrangian here does not imply that the Lagrangian introduced previously follows certain proper power counting rules, rather it means that the Lagrangian includes all the expected nearby (hence important) singularities of a given process, in the limited energy range of data fitting.

\subsection{ Calculation of the amplitudes }

\begin{figure}[htbp]
\includegraphics[width=0.53\textwidth]{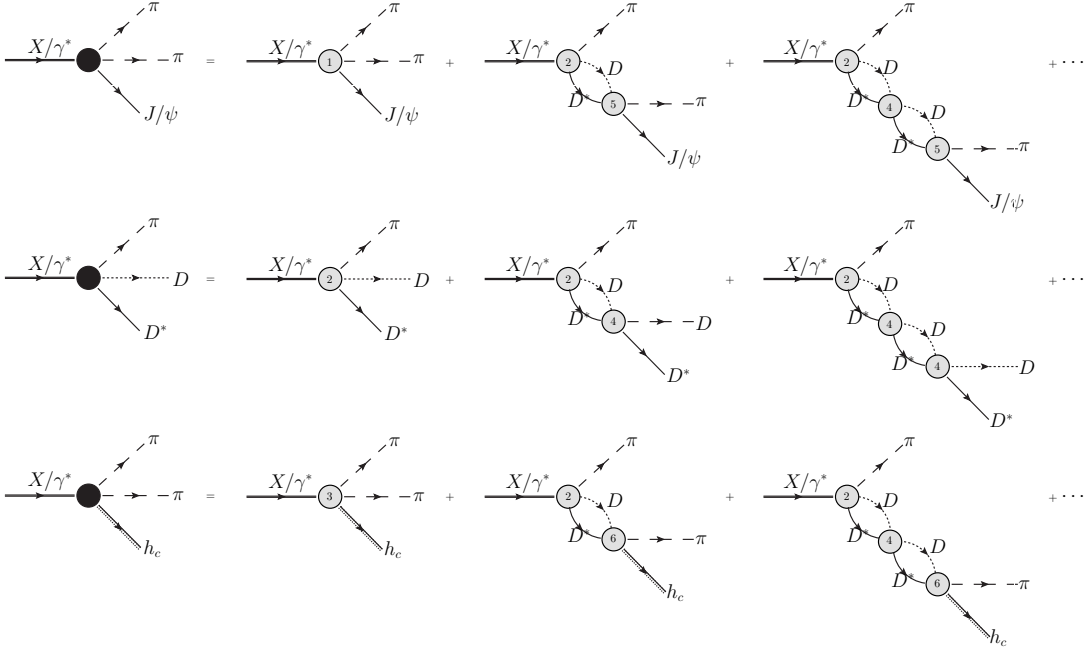}
\\
\vspace*{0.6cm}
\caption{Decay diagrams of $X(4260)$ or $\gamma^*$. The ellipses stand for the infinite series sum of the $\bar DD^*$ loops.  }\label{zc3}
\end{figure}

\begin{figure}[htbp]
\centering
  \includegraphics[width=0.75\textwidth]{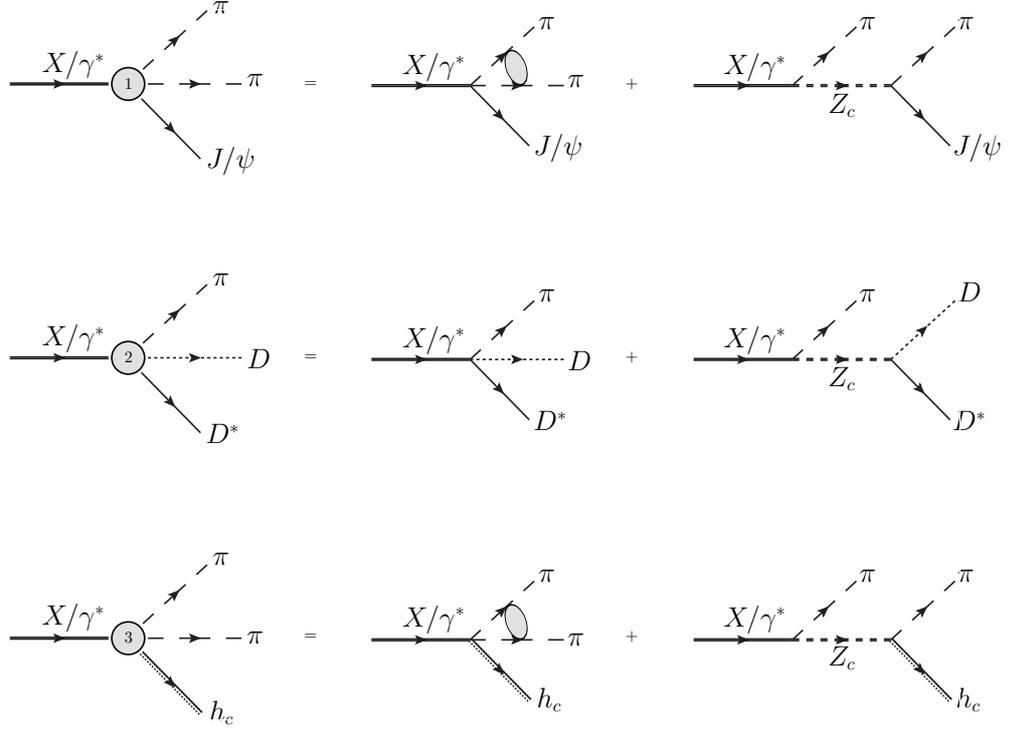}
  \caption{Decay composite vertices. }\label{zc1}
\end{figure}

\begin{figure}[htbp]
\centering
  \includegraphics[width=0.75\textwidth]{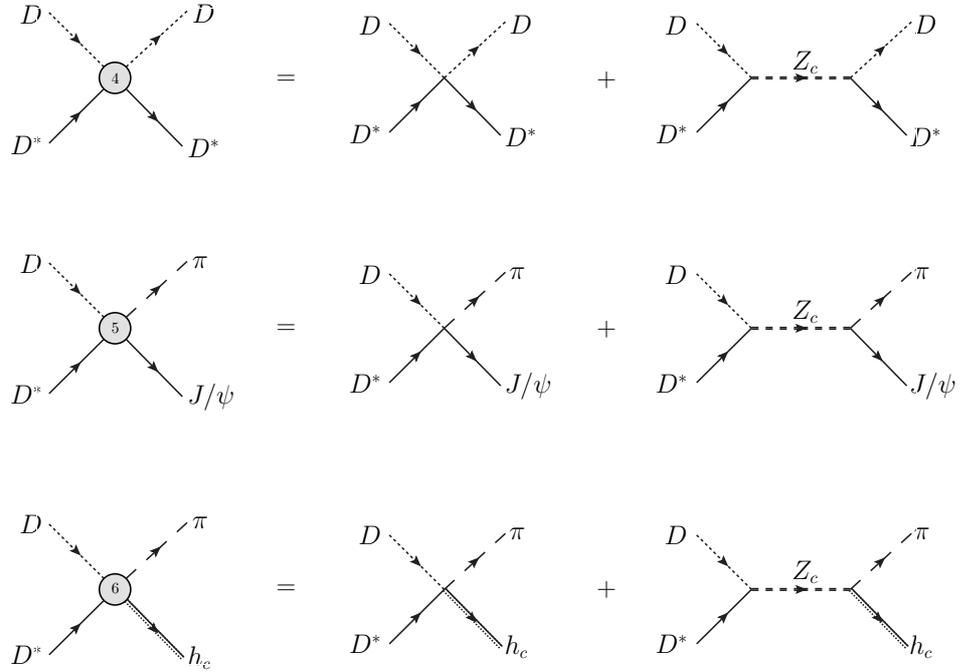}
  \caption{Four point composite vertices.}\label{zc2}
\end{figure}

Next we calculate the amplitudes corresponding to the $e^+e^-\to J/\psi\pi\pi$, $\bar{D}D^*\pi$ and $h_c\pi\pi$ processes. The center of mass (CM) energies
in those processes are fixed at 4.26~GeV in accord with the experimental analyses~\cite{Ablikim:2013mio,Ablikim:2013wzq,Ablikim:2013xfr}\footnote{ Note that three
different CM energies at 4.23~GeV, 4.26~GeV and 4.36~GeV are taken for the $h_c\pi\pi$ channel in Ref.~\cite{Ablikim:2013wzq}.
Nevertheless the impact on the mass distribution of $h_c\pi$ with the small variances of the CM energies is found to be negligible. Therefore we simply
fix the CM energy at 4.26~GeV for the $h_c\pi\pi$ channel, as done for the other two cases.}.
The final amplitudes generated by the effective Lagrangian in the previous subsection and also the important $\pi\pi$ FSIs are depicted separately for the $J/\psi\pi\pi, \bar{D}D^*\pi, h_c\pi\pi$ processes in Figs.~\ref{zc3}--\ref{zc2}. Let us explain the meaning of different symbols in the previous figures in more detail. The gray blob with a specific number indicates that this interacting vertex is a composite interaction, in the sense that it comprises more than one Feynman diagram dictated by the Lagrangian in the previous subsection.  The composite vertices are graphically defined in Figs.~\ref{zc1} and \ref{zc2}. The blobs labeled by $1,2,3$ represent the interactions between the initial state $X(4260)$ or the virtual $\gamma^*$ and the final states $J/\psi\pi\pi, \bar{D}D^*\pi, h_c\pi\pi$, without including any $\bar{D}D^*$ FSI. Their explicit definitions are given in Fig.~\ref{zc1}. Basically these three blobs describe the tree-level amplitudes of $e^+e^-\to J/\psi\pi\pi, \bar{D}D^*\pi, h_c\pi\pi$ by including the possible $\pi\pi$ FSIs\footnote{Strictly speaking, the blobs 1 and 3 are not tree-level amplitudes, since the infinite sum of the light-flavor meson loops are included through the  unitarization of the $\pi\pi$ scattering amplitude. }. The blobs labeled by $4,5,6$ stand for the tree-level transition amplitudes from $\bar{D}D^*$ to $\bar{D}D^*$, $J/\psi\pi$ and $h_c\pi$ respectively, which are explicitly shown in Fig.~\ref{zc2}.  As clearly depicted in Figs.~\ref{zc1} and \ref{zc2}, not only the contact interactions but also the bare $Z_c$ exchanges are explicitly included in our calculation.


The $\pi\pi$ FSI, denoted by the shaded circle with pion legs in Fig.~\ref{zc1}, is important only for the $IJ=00$ $\pi\pi$ system when they are produced from the contact vertex.
For the case when one of the pions is being produced by $Z_c$ emission, it should not be affected much by another pion previously produced, remembering that the lifetime of $Z_c$ is rather long compared with the typical hadron lifetime.
An explicit verification of this statement will be given in Sec.~\ref{sec.numeric}.
We closely follow Ref.~\cite{Dai:2012pb} to implement the $\pi\pi$ FSI. We simply give a sketch on how to include this effect here and refer to the previous reference for further details. The decay amplitude of , e.g., $X/\gamma^* \to J/\psi\pi\pi$ after including the $\pi\pi$ FSI takes the form
\begin{align}\label{fsipipi1}
&\mathcal{A}_1= \mathcal{A}_1^{tree} \alpha_1(s) T_{11}(s) + \mathcal{A}_2^{tree} \alpha_2(s) T_{21}(s) \,, \nonumber \\
&\mathcal{A}_2= \mathcal{A}_2^{tree} \alpha_1(s) T_{12}(s) + \mathcal{A}_2^{tree} \alpha_2(s) T_{22}(s) \,,
\end{align}
where the coupled channels with $\pi\pi$ (labeled as channel 1) and $\bar KK$ (labeled as channel 2) are considered. $\mathcal{A}_1$ stands for the expression of the diagram with the shaded circle between the pion legs in Fig.~\ref{zc1}.
$\mathcal{A}_1^{tree}$ ($\mathcal{A}_2^{tree}$) denotes the contact tree-level amplitude of $X/\gamma^* \to J/\psi\pi\pi\,\, (\bar KK)$ calculated using the Lagrangian Eq.~\eqref{lag}.
In Eq.~\eqref{fsipipi1} $\alpha_{1,2}(s)$ are mild polynomial functions. $T_{11}(s)$, $T_{12}(s)$(=$T_{21}(s)$) and $T_{22}(s)$ correspond to the unitarized isoscalar-scalar partial-wave amplitudes for $\pi\pi\to\pi\pi$, $\pi\pi\to\bar KK$ and $\bar KK \to \bar KK$, respectively~\cite{Dai:2011bs}. All of the unknown low energy constants in those unitarized amplitudes are fixed by fitting the scattering data.

With the preparation given  above, we are now ready to calculate the decay amplitudes of  $\pi \bar{D}D^*$, $J/\psi\pi\pi$ and $ h_c\pi\pi$ channels by including the strong $\bar DD^*$ FSI, i.e. the infinite series sum of the $\bar DD^*$ loops in Fig.~\ref{zc3}.
We make a careful study on the resummation of the infinite geometry series of $\bar DD^*$ loops by properly taking into account the general composite four-point vertex, which includes both the local contact interaction and also the $Z_c$ exchange, as depicted in the top row of Fig.~\ref{zc2}.
We demonstrate that in order to accomplish the resummation of $\bar DD^*$ loops one needs to split the amplitudes into two geometry series, i.e. the transverse part and the longitudinal one. In order not to interrupt the present discussion, we simply elaborate the essentials to obtain these results here, and the detailed calculations and explicit forms of the amplitudes after taking into account the $\bar DD^*$ bubble chains as depicted in Fig.~\ref{zc3}  are relegated to Appendix~\ref{appendix.amp}.

Notice that the composite rescattering vertex of $D\bar D^*$ is generated in two ways: contact interaction and $Z_c(3900)$ exchange in  $s$-channel as shown in the top row of Fig.~\ref{zc2}, and can be written as
\begin{equation}
  iA^{\mu\nu}=i\lambda_1g^{\mu\nu}-f_5^2D^{\mu\nu}(l^2)\ ,
\end{equation}
 where $l=p_{\bar D}+p_{D^*}$ is the momentum of $Z_c(3900)$, and its propagator $D^{\mu\nu}(l^2)$ reads
  \begin{align}
    D^{\mu\nu}(l^2)=-i\frac{g^{\mu\nu}-\frac{l^\mu l^\nu}{m^2_Z}}{l^2-m^2_Z},
  \end{align}
with $m_Z$ denoting the mass parameter of $Z_c(3900)$.
In order to sum up the infinite-loop chain, it is necessary to divide the propagator into transverse and longitudinal parts,
 \begin{equation}
  D^{\mu\nu}(l^2)=\frac{-i(g^{\mu\nu}-\frac{l^\mu l^\nu}{m^2_Z})}{l^2-m^2_Z}=\frac{-iP^{\mu\nu}_T(l^2)}{l^2-m^2_Z}+\frac{iP^{\mu\nu}_L(l^2)}{m^2_Z}\ ,
\end{equation}
with $P_{T}^{\mu\nu}=g^{\mu\nu}-\frac{l^\mu l^\nu}{l^2}$ and $P_{L}^{\mu\nu}=\frac{l^\mu l^\nu}{l^2}$. In Ref.~\cite{Meng:2014ota} a detailed analysis is made on the longitudinal part, concluding that the longitudinal part has only a  minor impact on the behavior of the amplitudes near $\bar{D}D^*$ threshold and serves as a background contribution only. Particularly it points out that the poles originated from the longitudinal amplitudes are far away from the $\bar DD^*$ threshold and hence are unphysical.

\section{Phenomenological analyses}\label{sec.numeric}
\subsection{ Description of the fit strategies }

After obtaining the full amplitudes of the $\pi \bar{D}D^*$, $J/\psi\pi\pi$ and $h_c\pi\pi$ channels, we make use of them to perform numerical fits in the following. We focus on two scenarios in the fits.
\begin{itemize}
\item Fit I: Consider only the $(\bar D D^*)^2$ contact interaction. This is realized by switching off all $Z_c$ couplings to other particles, that is to fix the couplings $g_4, g_5, f_5$ and $f_7$ in Eqs.~(\ref{eq2a},~\ref{eq2b},~\ref{eq2c},~\ref{eq2d}) to zero. In this situation we test the molecular nature of $Z_c(3900)$.

\item Fit II: Assume that there indeed exists a bare $Z_c(3900)$ state, which is described by a Flatt\'e propagator.\footnote{An explicit resonance exchange with the Flatt\'e-type parametrization does not automatically guarantee an elementary state, rather it may still simulate a molecular state according to the pole counting rule. We will elaborate  this point in detail later. } Meanwhile we turn off the $(\bar D D^*)^2$ four-point contact interaction vertex, that is to fix the coupling $\lambda_1$ in Eq.~\eqref{eq1a} to zero.
\end{itemize}
Of course, in addition to the above two fits we also test the mixed situation, that is to include both the $(\bar DD^*)^2$ four-point contact interaction and the bare $Z_c$ exchange. We will briefly discuss the mixed-type fit result in the following subsection.

In Fit I, there is only a $(\bar{D}D^*)^2$ contact interaction and the $Z_c(3900)$-exchange contributions are excluded. To be more specific, we fix
the couplings $g_4, g_5, f_5$ and $f_7$ in Eqs.~(\ref{eq2a},~\ref{eq2b},~\ref{eq2c},~\ref{eq2d}) to zero in this case. Then the denominators of transverse and longitudinal amplitudes presented in the Appendix~\ref{appendix.amp} (e.g. Eq.~\eqref{eqjpsiaa} ) take the form
\begin{equation}
        1-i\lambda_{1}\Pi_{T}\,, \qquad ~1-i\lambda_{1}\Pi_{L}\, , \label{eq3}
\end{equation}
respectively, with $\lambda_1$ characterizing the strength of $(\bar{D}D^*)^2$ contact interaction, cf. Eq.~\eqref{eq1a}. The functions  $\Pi_{T}$ and $\Pi_{L}$ are defined in Eqs.~\eqref{pit}) and \eqref{pil}, respectively.  However, in reality the $\bar DD^*$ may scatter into other lighter channels, and this fact is taken into account by introducing a constant parameter in Eq.~(\ref{eq3}):
\begin{equation}\label{eq14}
        1-i\lambda_{1}(\Pi_{T}+c_0)\,\qquad ~1-i\lambda_{1}(\Pi_{L}+c_0)\,,
\end{equation}
where $c_0$ is real and accounts for the effects of the channels that are far below the $\bar DD^*$ threshold. Its major effect is to bring a possible decay width (into light channels) to the $\bar DD^*$ molecule.

In Fit II, $\bar{D}D^*$ interacts only through exchanging intermediate $s$--channel $Z_c(3900)$. There is no $\bar{D}D^*$ contact interaction and $\lambda_1$ is fixed to zero.  At the same time, the coupling parameters in Eqs.~\eqref{eq1b} and \eqref{eq1c} are also set to zero, implying that we do not consider the contact interactions between $\bar DD^*$ and $J/\psi\pi, h_c\pi$ in Fit II.  Then the denominators of transverse and longitudinal amplitudes presented in the Appendix~\ref{appendix.amp} (e.g. Eq.~\eqref{eqjpsiaa} ) take the following form
       \begin{equation}\label{eq15}
       1-\frac{if_{5}^{2}}{l^{2}-m_{Z}^{2}}\Pi_{T}\, \qquad ~1+\frac{if_{5}^{2}}{m_{Z}^{2}}\Pi_{L}\,,
      \end{equation}
respectively, where $f_5$ represents the coupling strength between $Z_c(3900)$ and $\bar{D}D^*$, cf. Eq.~\eqref{eq2c}. Apart from the $\bar DD^*$, $J/\psi\pi$ and $h_c\pi$ channels, there could be other decay channels whose thresholds are much lighter than the production energy of $\zc$. Since all such  thresholds are far away from the $\bar DD^*$ one, their contributions to the $Z_c(3900)$ decay width are approximately parameterized as a constant $\Gamma_0$.
In order to account for these contributions, the denominator of the transverse propagator in Eq.~(\ref{eq15}) is replaced  by
\begin{equation}\label{eq1}
       l^{2}-m_{Z}^{2}+i m_Z \Gamma_{Z_c}(l^2),
      \end{equation}
where $\Gamma_{Z_c}(l^2)=\Gamma_{J/\psi\pi}(l^2)+\Gamma_{h_c\pi}(l^2)+\Gamma_0$, with $\Gamma_{J/\psi\pi}(l^2)$ and $\Gamma_{h_c\pi}(l^2)$ the partial widths of the $Z_c(3900)$ for corresponding channels.
 The widths $\Gamma_{J/\psi\pi}(l^2)$ and $\Gamma_{h_c\pi}(l^2)$ are given, respectively, by
 \begin{align}
   \Gamma_{J/\psi\pi}(l^2)=\frac{|p_{\psi}|}{8\pi l^2}|M_{J/\psi\pi}|^2,
 \end{align}
 \begin{align}
   \Gamma_{h_c\pi}(l^2)=\frac{|p_{H}|}{8\pi l^2}|M_{h_c\pi}|^2,
 \end{align}
where $|p_{\psi}|=\sqrt{\frac{(l^2-(m_{J/\psi}+m_{\pi})^2)(l^2-(m_{J/\psi}-m_{\pi})^2)}{4l^2}}$ and $|p_{H}|=\sqrt{\frac{(l^2-(m_{h_c}+m_{\pi})^2)(l^2-(m_{h_c}-m_{\pi})^2)}{4l^2}}$; $m_{J/\psi}$, $m_{h_c}$ and $m_{\pi}$ are the mass of $J/\psi$,
$h_c$ and $\pi$, respectively. In addition,  $M_{J/\psi\pi}$ and $M_{h_c\pi}$ are the tree-level amplitudes of $Z_c(3900)$ decaying  into $J/\psi\pi$ and $h_c\pi$, which can be calculated using the Lagrangians in Eqs.~\eqref{eq2b} and \eqref{eq2d},  respectively,
\begin{align}
  M_{J/\psi\pi}=-i g_5 p_\pi\cdot p_\psi \epsilon_Z\cdot \epsilon_\psi,
\end{align}
\begin{align}
  M_{h_c\pi}=-i f_7p_{H}^\alpha p_{\pi}^\beta \epsilon^{\mu\nu\alpha\beta}\epsilon_Z^\mu\epsilon_{H}^\nu,
\end{align}
where $p_\psi$, $p_{H}$ and $p_\pi$ are the momenta of $J/\psi$, $h_c$ and $\pi$; $\epsilon_Z$, $\epsilon_\psi$ and $\epsilon_H$ are the polarization vectors of $Z_c(3900)$, $J/\psi$ and $h_c$, respectively.

Before ending this section, we briefly comment the difference between our approach and the one in Refs.~\cite{Hanhart:2015cua,Guo:2016bjq}. The latter two references used Lippmann-Schwinger equations to properly take into account the coupled-channel unitarities to study the $Z_b$ spectrum, while in our approach only the elastic unitarity in $\bar{D}D^{*}$ channel is strictly incorporated through the bubble-chain resummation, since this is the most important channel responsible for the $Z_c(3900)$ peak. For the inelastic channels, such as $J/\psi \pi$ and $h_c \pi$, we make an approximation  to consider their unitarization effects, that is to introduce their finite widths in the denominator of the $Z_c$ propagator, cf. Eq.~\eqref{eq1}. One should not expect strong nonperturbative effects appearing among these inelastic channels. In other words, the unitarizations of these two channels are unlikely to  produce any new significant singularity. Furthermore we focus on the energy region around $3900$~MeV in the present work, while the $J/\psi\pi$ and $h_c\pi$ thresholds are more than 600 and 200~MeV below the $Z_c(3900)$ peak, respectively. Therefore our way to include the unitarization effects from the $J/\psi \pi$ and $h_c \pi$ channels should be justified around the $Z_c(3900)$ region.

\subsection{ Data Fitting and Numerical Results}\label{sec.fitres}

We make a combined fit to the data on the $J/\psi\pi^\pm$ and $\pi\pi$ invariant mass spectra in the $J/\psi\pi\pi$ channel~\cite{Ablikim:2013mio,Liu:2013dau,Lees:2012cn}, $\bar{D}D^*$ mass distributions in the $\bar{D}D^*\pi^{\pm}$ channel~\cite{Ablikim:2013xfr} and $h_c\pi^\pm$ invariant mass spectrum in $h_c\pi^{+}\pi^{-}$ channel~\cite{Ablikim:2013wzq}.
The energy resolution of different channels is also considered. To be more specific, the $J/\psi\pi\pi$ amplitude that has been projected to $s$ wave of the $\pi\pi$ system by including also the $\pi\pi$ FSI~\cite{Dai:2012pb,Dai:2011bs}, is convolved with a Gaussian function with the energy resolution fixed to be $\sigma=4.2\mbox{MeV}$~\cite{Ablikim:2013mio}, which can be written as follows: 
\begin{align}\label{engresol}
  \Gamma(l)=\int_{l-3\sigma}^{l+3\sigma}dl'\frac{1}{\sqrt{2\pi}\sigma}\Gamma(l')\exp^{-\frac{(l'-l)^2}{2\sigma^2}}.
\end{align}
Here $l=m_{J/\psi\pi}$ is the momentum of $Z_c(3900)$ and $\Gamma$ is the cross section of $Y(4260)/\gamma^*\to J/\psi\pi\pi$.\footnote{However, the final result does not rely much on whether the energy resolution is considered or not.} On the other side, the energy resolutions of
$ h_c\pi\pi$ and $\bar{D}D^*\pi$ channels are $1.8~\mbox{MeV}$~\cite{Ablikim:2013wzq} and $1~\mbox{MeV}$~\cite{Ablikim:2013xfr} respectively, which can be safely ignored.

We point out that the $\pi\pi$ spectrum in the $J/\psi\pi\pi$ channel has been carefully studied, while this is not the case in the recent works in Refs.~\cite{Guo:2014iya,Albaladejo:2015lob}. Although the $\pi\pi$ dynamics gives more like a background contribution to the interested $\zc$ energy region, its coherent interference with other terms in the full amplitude can provide non-negligible effects. Three sets of $\pi\pi$ data in $J/\psi\pi^{+}\pi^{-}$ channel are fitted, which are
from BESIII~\cite{Ablikim:2013mio}, Belle~\cite{Liu:2013dau} and BaBar~\cite{Lees:2012cn}. Besides $\pi\pi$ data,
we also fit the $\bar{D}D^*$  (including $D^{-}D^{*0}$ and $D^{+}\bar{D}^{*0}$) data from $3.87\mbox{GeV}$ to $4.11\mbox{GeV}$ of Ref.~\cite{Ablikim:2013xfr}, the {$M_{max}(J/\psi\pi^{\pm})$} data from $3.67\mbox{GeV}$ to $4.1\mbox{GeV}$ of Ref.~\cite{Ablikim:2013mio}, \footnote{The maximum spectrum $M_{max}(J/\psi\pi^{\pm})$ is the mass distribution of the larger one of $M_{J/\psi\pi^+}$ and $M_{J/\psi\pi^-}$ which only shows one $Z_c$ peak but gives equivalent information, compared with the $M_{J/\psi\pi^\pm}$ mass distribution~\cite{Ablikim:2013mio}.}and the $h_c\pi^\pm$ data from $3.80\mbox{GeV}$ to $3.93\mbox{GeV}$ in Ref.~\cite{Ablikim:2013wzq}.

\begin{figure}[htbp]
\center
\subfigure[]{
\label{fig:subfig:a}
\scalebox{1.2}[1.2]{\includegraphics[height=1.5in,width=2.3in]{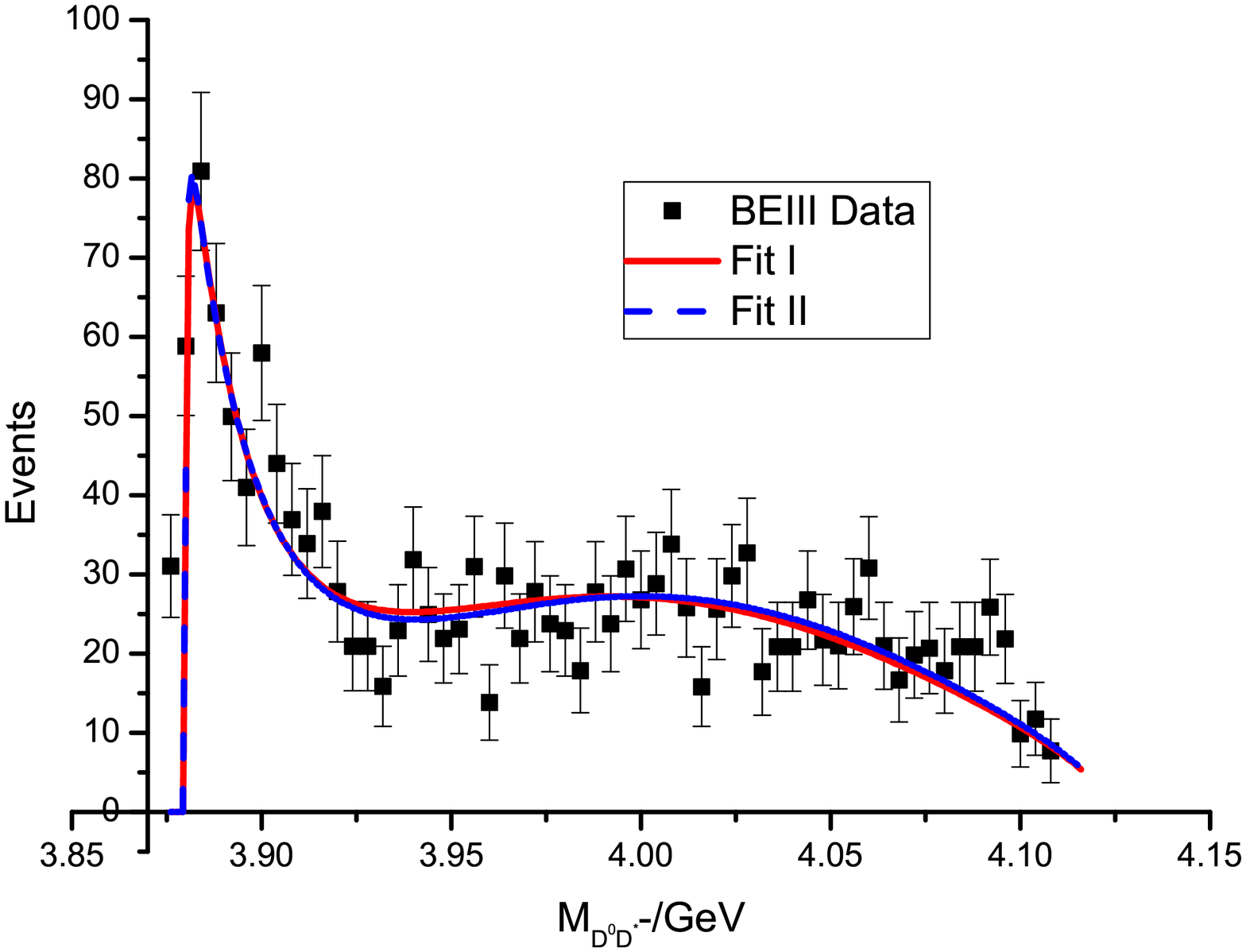}}}
\subfigure[]{
\label{fig:subfig:b}
\scalebox{1.2}[1.2]{\includegraphics[height=1.5in,width=2.3in]{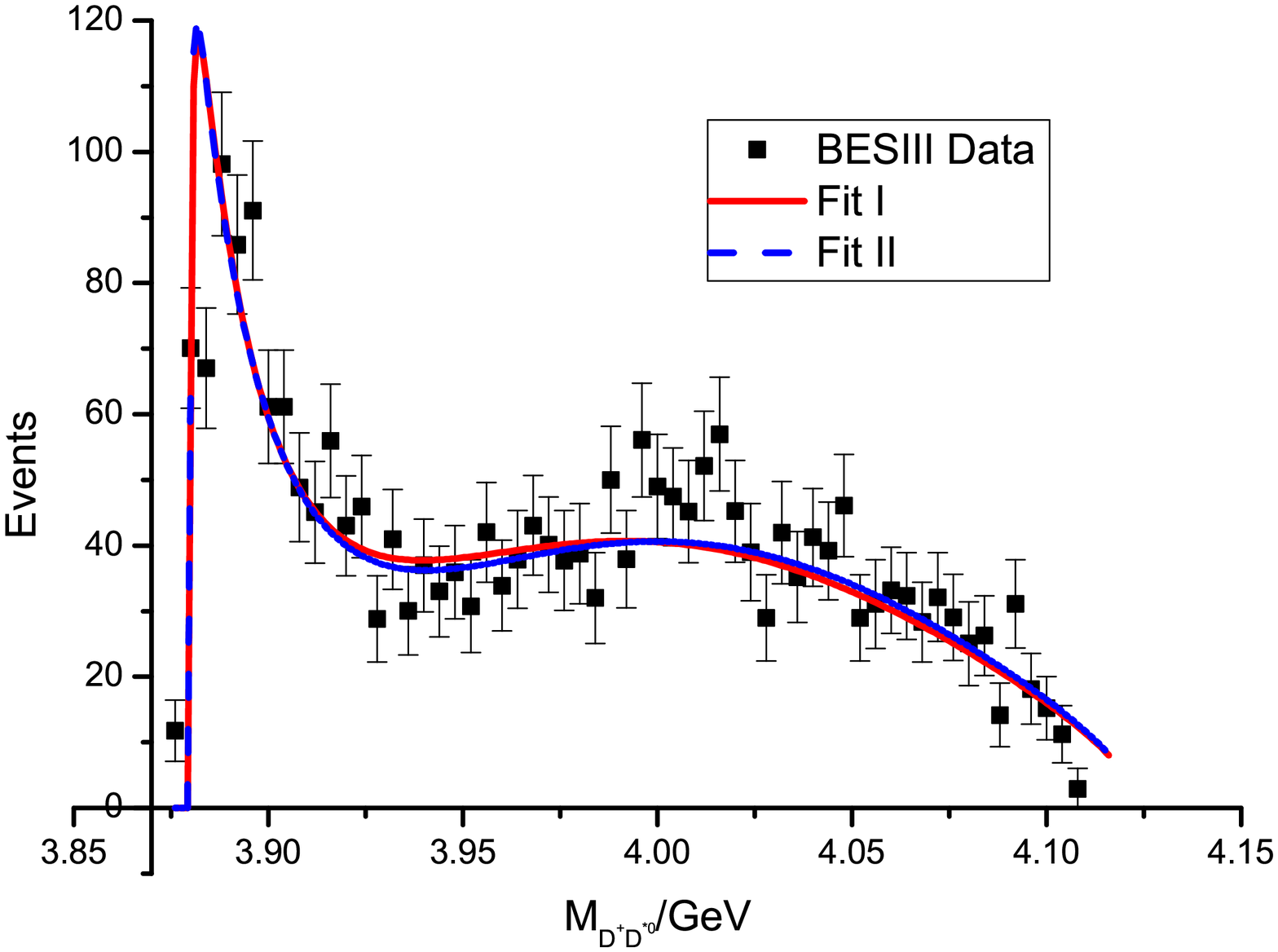}}}\\
\subfigure[]{
\label{fig:subfig:c}
\scalebox{1.1}[1.1]{\includegraphics[height=1.5in,width=2.4in]{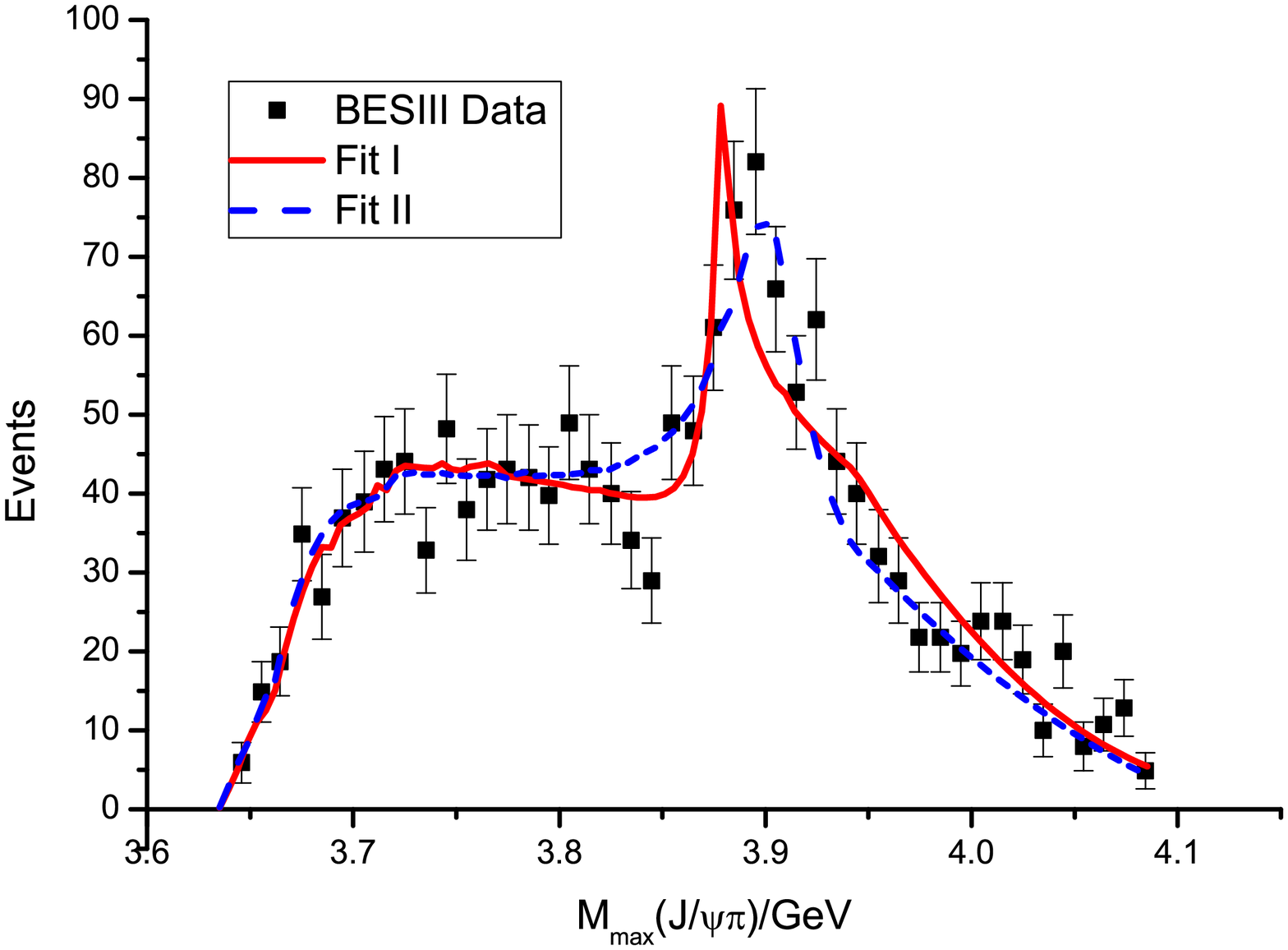}}}
\subfigure[]{
\label{fig:subfig:d}
\scalebox{1.1}[1.1]{\includegraphics[height=1.5in,width=2.4in]{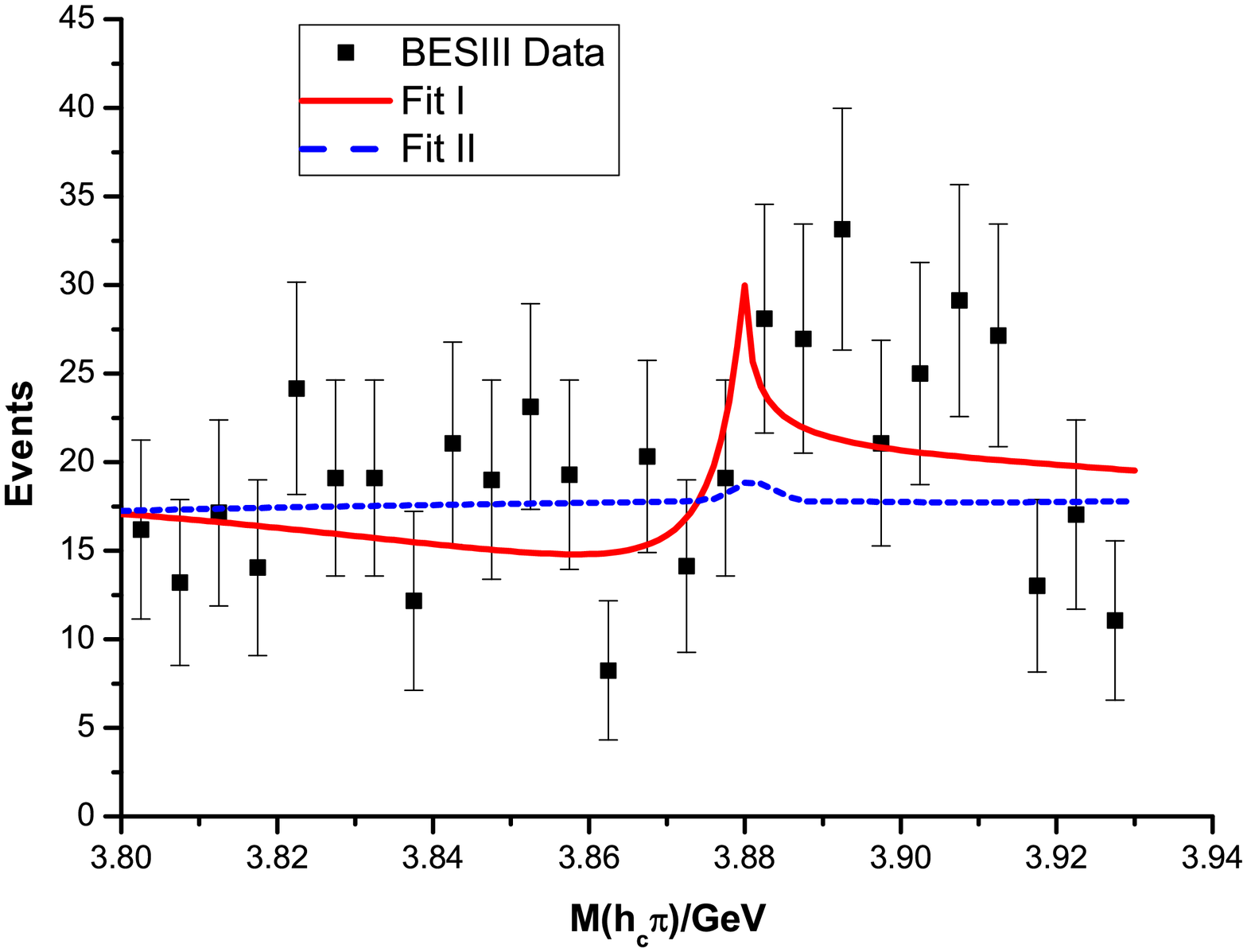}}}
\caption{
 (a): $D^0D^{*-}$ invariant mass spectrum from Ref.~\cite{Ablikim:2013xfr}; (b): $D^+\bar{D}^{*0}$ invariant mass spectrum from Ref.~\cite{Ablikim:2013xfr};
 (c): $J/\psi\pi^{\pm}$ maximum invariant mass spectrum from Ref.~\cite{Ablikim:2013mio}; (d): $h_c\pi^{\pm}$ invariant mass spectrum from Ref.~\cite{Ablikim:2013wzq}.}\label{fig5}
\end{figure}

\begin{figure}[htbp]
\center
\subfigure[]{
\label{fig:subfig:4a}
\scalebox{1.2}[1.2]{\includegraphics[height=1.5in,width=2.3in]{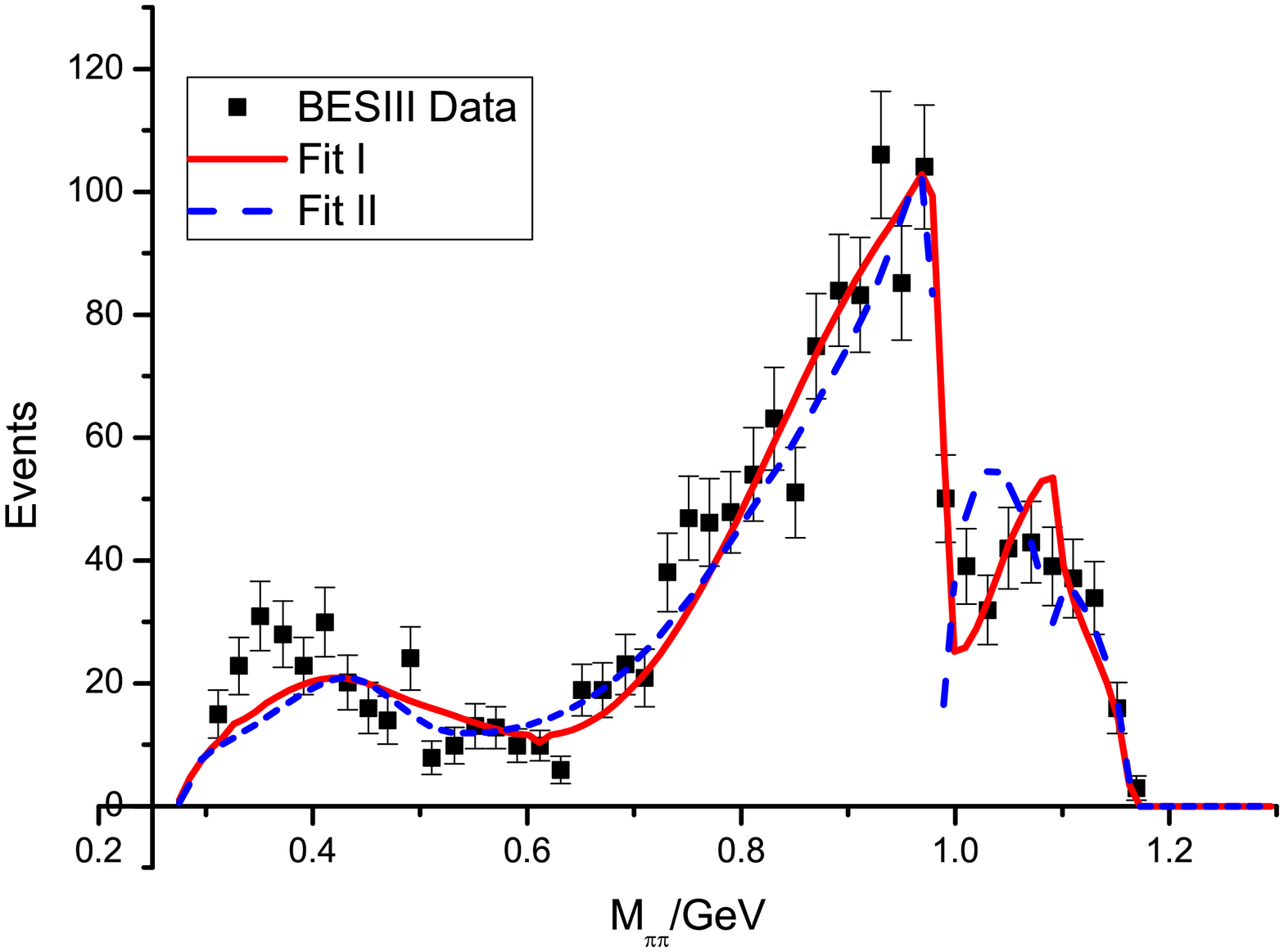}}}
\subfigure[]{
\label{fig:subfig:4b}
\scalebox{1.2}[1.2]{\includegraphics[height=1.5in,width=2.3in]{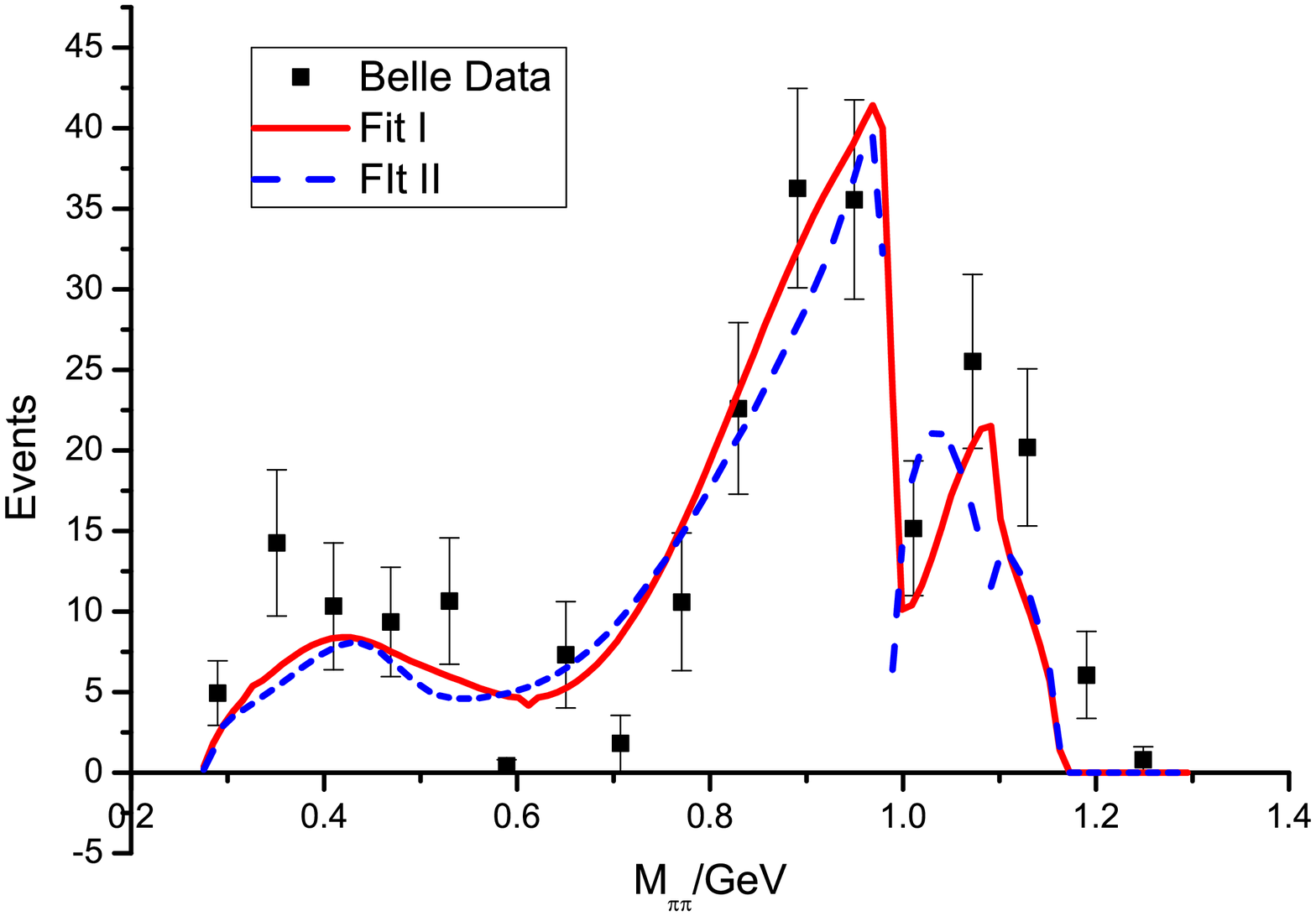}}}
\subfigure[]{
\label{fig:subfig:4c}
\scalebox{1.2}[1.2]{\includegraphics[height=1.5in,width=2.5in]{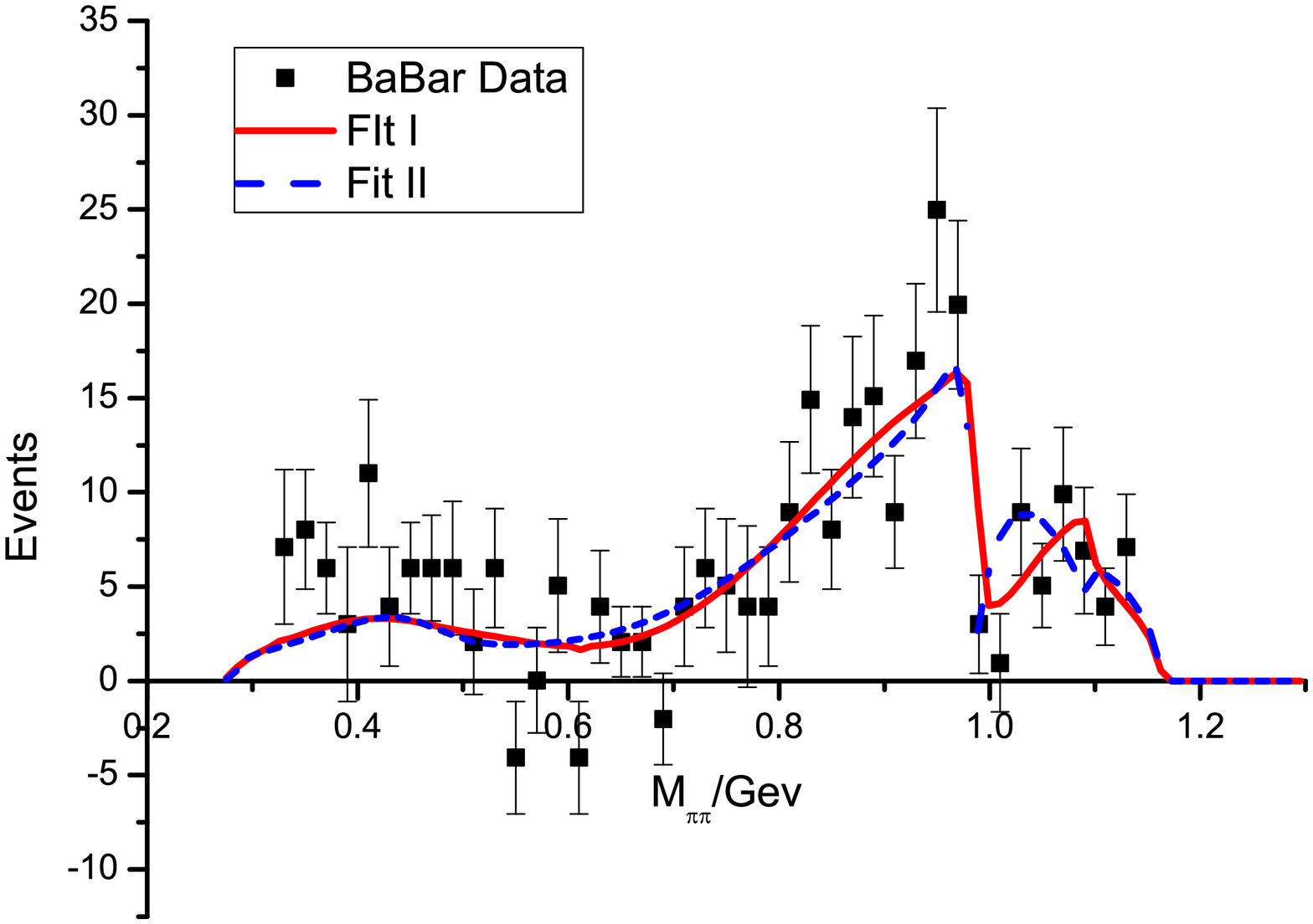}}}
\caption{  The $\pi\pi$ invariant mass spectra of the $e^{+}e^{-}\rightarrow J/\psi\pi\pi$ process. (a), (b) and (c) correspond to $\pi\pi$ invariant mass spectra from Ref.~\cite{Ablikim:2013mio}, Ref.~\cite{Liu:2013dau} and Ref.~\cite{Lees:2012cn}, respectively.}\label{fig4}
\end{figure}

There are 16 and 14 coupling parameters for Fit I and Fit II, respectively. Adding four parameters of $\pi\pi$ final state interaction~\cite{Dai:2012pb}, seven normalization parameters and one parameter for $D\bar{D}^*$ (incoherent) background, the number of  parameters in total is 28 and 26 for Fit I and Fit II, respectively.
The fit results are plotted in Figs.~\ref{fig5} and ~\ref{fig4}.

It is found that the $\chi^2/d.o.f=497/(291-26)$ in Fit II is a little bit larger than $\chi^2/d.o.f=454/(291-28)$ of Fit I. \footnote{If we drop out $\pi\pi$ data and only include  $DD^*$, $J/\psi\pi$ and
$h_c\pi$ data, the $\chi^2/d.o.f$ would be  $282/(189-21)$ in Fit I and  $274/(189-19)$ in Fit II. }
From our experience in data fitting, this result is very interesting -- naively, one might expect that the bubble-chain description would not fit the data very well comparing with the standard concise Flatt\'e description with explicit resonance exchange.
A quick conclusion might be that this result does not seem to disfavor the molecule picture comparing the elementary $Z_c$  mechanism. We will in fact realize soon in the next subsection, that the Flatt\'e description for the explicit resonance exchange in the present analysis is actually ``dynamically'' equivalent to the bubble chain parametrization, according to the pole counting rule.

Though the overall quality of Fit I and Fit II are quite similar, there are some minor differences between the two fits. For the $\bar{D}D^*$ invariant mass spectrum, the two fits both agree well with the data. The peak near $3900\mbox{MeV} $ in  $J/\psi\pi$ spectrum in Fit II is a bit wider than that in Fit I, while for $h_c\pi^\pm$ spectrum, the molecule mechanism produces an enhancement near $3.88\mbox{GeV}$ in Fig.~\ref{fig:subfig:d}, which is, however, absent in the bare $Z_c$-exchange picture.

In principle, the most general fit is the mixed one by including both $(\bar DD^*)^2$ contact interactions and $Z_c(3900)$ exchanges as shown in Figs.~\ref{zc1} and \ref{zc2}. In this case, the denominators of transverse and longitudinal amplitudes presented in Appendix~\ref{appendix.amp} (e.g. Eq.~\eqref{eqjpsiaa} ) take the form
\begin{align}
  1-i(\lambda_{1}+\frac{f_{5}^{2}}{l^{2}-m_{Z}^{2}})\Pi_{T}\,,\qquad ~1-i(\lambda_{1}-\frac{f_{5}^{2}}{m_{Z}^{2}})\Pi_{L}\,.
\end{align}
However, we explicitly verify that this way to perform the fits does not obviously improve the total $\chi^2$ comparing with Fit I and Fit II. Since more parameters are involved in the ``mixed" mechanism, the fit procedure becomes more unstable compared with Fit I and Fit II. Therefore, at this level of study no useful information can be extracted from the ``mixed'' fit, and we refrain from discussing it further.

\begin{figure}[htbp]
\centering
\includegraphics[width=0.7\textwidth]{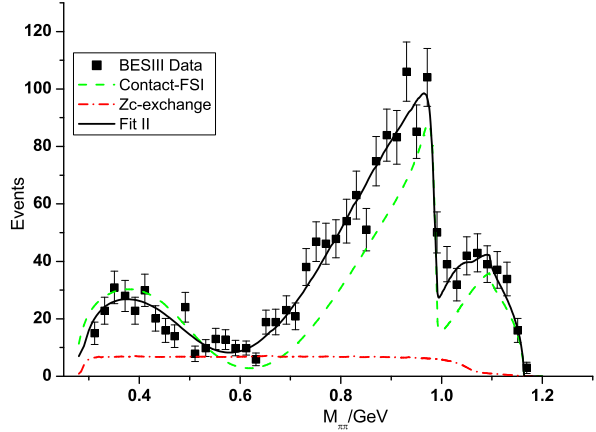}
\caption{The black solid line denotes the $\pi\pi$ invariant mass distribution from Fit II. The green dashed line denotes the contribution from the pure contact amplitude with $\pi\pi$ FSI. The red dashed-dotted line represents the $Z_c$-exchange contribution.   }\label{fig.pipi}
\end{figure}

\begin{figure}[htbp]
\centering
\includegraphics[width=0.7\textwidth]{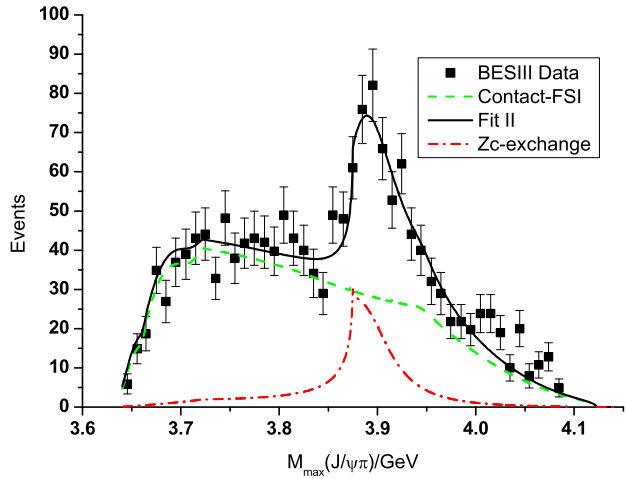}
\caption{The black solid line denotes the $J/\psi \pi$ invariant mass distribution from Fit II. The other notations are the same as those in Fig.~\ref{fig.pipi}. } \label{fig.jpsipi}
\end{figure}

Now it is in order to comment on the relative strengths between different mechanisms in the $\pi\pi$ and $J/\psi \pi$ spectra. In Fig.~\ref{fig.pipi}, we show different contributions from the last two diagrams in the first line of Fig.~\ref{zc1}, i.e. the dressed contact vertex with $\pi\pi$ FSI and the tree-level $Z_c$ exchange. It is clear that the former contribution dominates the $\pi\pi$ invariant mass distribution and the latter behaves more like a small background term. The anatomy of different contributions in $J/\psi \pi$ spectrum is given in Fig.~\ref{fig.jpsipi}.  Though the peak structure around 3.9~GeV is mainly contributed by the $Z_c$ exchange diagram, the $\pi\pi$ FSI dressed contact vertex gives important background effects.

\subsection{ Pole analysis}\label{sec.pole}

According to the pole counting rule~\cite{Morgan:1992ge}, a molecule generated in $s$-wave scattering  can be distinguished from an elementary particle by counting the number of poles near the relevant physical threshold on different Riemann sheets. In this subsection, we search for poles of the previously determined amplitudes in the complex energy plane.

Three thresholds are relevant, namely $J/\psi\pi$, $h_c\pi$ and $\bar{D}D^*$.
However, through the fit procedure it is found that the $h_c\pi$ channel plays only a minor role, i.e., it has only a negligible partial width. Hence as a good approximation the coupled-channel system reduces to a two-channel problem. We define in Table.~\ref{table1} the naming scheme of different Riemann sheets. From Eqs.~(\ref{eqjpsiaa}--\ref{hc}), one can see that the relevant poles in our amplitudes correspond to the zeros of the transverse denominator. As commented previously, the poles resulting from the longitudinal part are far away from the focused energy region and unphysical~\cite{Meng:2014ota}.
We therefore search for poles in Eqs.~(\ref{eq14}) and (\ref{eq1}) on four Riemann sheets characterized by the $\bar{D}D^*$ and $J/\psi\,\pi$ thresholds.
 \begin{table}[ht]
\begin{center}
 \begin{tabular}  {||c| c c c c||}
 \hline
   & sheet I & sheet II & sheet III & sheet IV \\
   \hline
  $\rho_{J/\psi\pi}(s)$ &$ +$ & $-$ & $-$ & $+$ \\
  \hline
  $\rho_{\bar{D}D^*}(s)$ &$ +$ & $+$ & $-$ & $-$ \\
  \hline
 \end{tabular}\\
 \caption{ The sign of the kinematic factors $\rho(s)$ in Eq.~\eqref{defrho} defines four Riemann sheets.}\label{table1}
   \end{center}
\end{table}
In Table~\ref{table2} we list all the nearby poles around the $\bar DD^*$ threshold. If the $h_c\pi$ data are not included in Fit I, then the pole will be located on sheet IV.\footnote{
The pole location here is controlled by the sign of  $\lambda_1$ parameter (the contact coupling of $\bar{D}D^*\bar{D}D^*$). If $\lambda_1$ is positive,  the pole is in sheet II, which means $Z_c$ is a molecular bound state of $\bar{D}D^*$; if $\lambda_1$ is
negative, a pole in sheet IV would be found which means $Z_c(3900)$ is a virtual state of $\bar{D}D^*$.  } On the contrary, the (nearby) pole always locates on sheet IV for Fit II.
The major difference between an explicitly introduced resonance and the molecule generated by a bubble-chain mechanism is that there are always intrinsic $two$ (pair of) poles  built in the former case in the coupled-channel situation.  But our numerical analysis shows that in Fit II,  one of the two poles in the Flatt\'e propagator is far away from the $\bar DD^*$ threshold and hence is not physically relevant at all.\footnote{ {The distant  pole locates at
$2.88\pm 0.35 i~\mbox{GeV}$ in sheet III. This pole is not actually meaningful any more since it is far outside the energy region under  control.}} Therefore we reach the most important physical conclusion in this work: Although the Flatt\'e-type parametrization and the bubble-chain mechanism look very different from the beginning,  they are practically equivalent in the present case. This observation  confirms the molecule nature of $Z_c(3900)$ according to the pole counting rule~\cite{Morgan:1992ge}.\footnote{In Ref.~\cite{Baru:2003qq}, the spectral-density-function (SDF) approach is employed to calculate the elementariness and compositeness coefficients. In Appendix \ref{sec.sdf}, we follow Refs.~\cite{Baru:2003qq,kala:2009na} to use the SDF method to calculate the elementariness coefficient of $Z_c(3900)$. Though suffering from large uncertainties, it seems to indicate that $\bar DD^*$ is the dominant component inside $Z_c(3900)$, which confirms the pole counting rule result.}  Here we would like to emphasize that ``close to the threshold"  does not in any sense necessarily lead to a molecular picture. A good counterexample is the $X(3872)$ resonance, it is also very close to the $\bar DD^*$ threshold. Nevertheless, an application of the pole counting rule indicates that it is mainly of $\bar cc$ nature~\cite{Meng:2014ota,Zhang:2009bv}. This conclusion agrees with results from other approaches~\cite{Suzuki:2005ha,Meng:2005er, kala}.
\begin{table}[ht]
\begin{center}
 \begin{tabular}  {||c| c c c c||}
 \hline
    & sheet I & sheet II & sheet III & sheet IV \\
   \hline
 Fit I &$ -$ & $3.87988\pm 0.00390 i$ & $-$ & $-$ \\
  \hline
  Fit II&$-$ & $-$ & $-$ & $3.87909\pm 0.00143i$ \\
  \hline
 \end{tabular}\\
 \caption{ The  pole locations from  Fit I and Fit II, with $h_c\pi$ data included. The numbers are given in units of $\mbox{GeV}$. }\label{table2}
   \end{center}
\end{table}

\begin{figure}[htbp]
 \centering
  \includegraphics[width=0.5\textwidth]{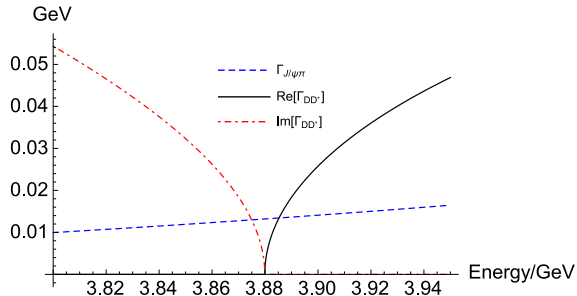}
  \caption{The energy dependence of the partial decay  width of $Z_c(3900)$. }\label{gammaaa}
\end{figure}

For completeness we also show in Table~\ref{table3} all partial widths from Fit II. Notice that here we do not include in the fit the constant width $\Gamma_0$ in Eq.~(\ref{eq1}). The reason is that since the $J/\psi\pi$ threshold is also quite low from the $\bar DD^*$ threshold, its $q^2$ dependence is weak, making it hard to distinguish from the constant $\Gamma_0$ (see Fig.~\ref{gammaaa} for an illustration). In other words, the value of $\Gamma_{J/\psi\pi}$ given in Table~\ref{table3} should  be better understood as a sum of $\Gamma_{J/\psi\pi}$ and $\Gamma_0$. Besides, the partial width $\Gamma_{\bar{D}D^*}$ is obtained  using the tree-level
decay width formula and the mass is chosen at the peak position (the line-shape mass), since the pole is slightly below the $\bar{D}D^*$ threshold. In Fig.~\ref{gammaaa}, we show explicitly
the strong energy dependences of $\Gamma_{Z_c\to\bar{D}D^*}$ and the almost flat behavior of $\Gamma_{Z_c\to J/\psi\pi}$.
An important conclusion from this plot is that the ratio of $\Gamma_{Z_c\to\bar{D}D^*}/\Gamma_{Z_c\to J/\psi\pi}$ is quite
sensitive to the pole position of $Z_c(3900)$ and bears a large uncertainty.

On the other hand, in the parametrization of Fit I, the possible decay of $D\bar D^*$ molecule into light channels is taken into account by adding  an additional parameter $c_0$ as done in Eq.~(\ref{eq14}), which is, however, difficult to directly connect  to partial widths.

\begin{table}[ht]
\begin{center}
\begin{tabular} {||c|c c c||}
\hline
Partial decay width& $\Gamma_{J/\psi\pi}$ &  $\Gamma_{\bar{D}D^*}$ & $\Gamma_{h_c\pi}$ \\
\hline
Value in \mbox{MeV}&$13.33\pm 0.62$&$16.47\pm 0.72$&$0.04\pm 0.03$\\
\hline
\end{tabular}\\
\caption{The resulting partial decay widths of $Z_c(3900)$ from Fit II. }\label{table3}
\end{center}
\end{table}

At the end of this section, we should mention that, in the above fits, the $h_c\pi$ data are included. Nevertheless,
since no significant $Z_c(3900)$ signal is observed in the $h_c\pi$ spectrum~\cite{Ablikim:2013wzq}, we also perform other fits by assuming that $Z_c(3900)$ does not decay into $h_c\pi$ and do not include the $h_c\pi$ data. The output is rather interesting: For Fit I we find that the pole  moves from sheet II to sheet IV when excluding the  $h_c\pi$ data. It implies that $Z_c(3900)$ is a bound state of $\bar{D}D^*$ when including the $h_c\pi$ data in the fit, and becomes a virtual state when $h_c\pi$ data are excluded. On the contrary, for Fit II the pole will always reside on sheet IV. It is worthy pointing out that the conclusion of the molecular nature of $Z_c(3900)$ is not changed regardless of including or excluding the $h_c\pi$ data in the fits.

\section{Discussions and Conclusions}\label{sec.conclu}

This work is devoted to the study of the nature of the $Z_c(3900)$ state. For this purpose we construct relevant effective Lagrangians which incorporate all possibly important singularities close to the $\bar DD^*$ threshold to calculate the $e^+e^-\to J/\psi\pi\pi, h_c\pi\pi$ and $\bar DD^*\pi$ processes. The $\pi\pi$ final state interactions are included for the $J/\psi\pi\pi, h_c\pi\pi$ channels by using the unitarized chiral perturbation theory, while for the strong $\bar DD^*$ interaction, we carefully perform
the infinite series sum of the $\bar DD^*$ loops. Hence we provide a good parametrization form to fit the relevant data, which include the $J/\psi\pi$ and $\pi\pi$ distributions from $e^+e^-\to J/\psi\pi\pi$, the $h_c\pi$ invariant mass spectrum from $e^+e^-\to h_c\pi\pi$ and the $\bar D D^*$ spectrum from $e^+e^- \to \bar DD^*\pi$.

Two different fits are performed with quite different physical motivations: Fit I, which includes only the $(\bar DD^*)^2$ contact interactions, is responsible for the examination of the molecular mechanism, and Fit II, which includes only the Flatt\'e form of the $Z_c(3900)$ exchange, is to test the elementary picture. It is remarkable that the two seemingly very different approaches point to the same conclusion: $Z_c(3900)$ is a $\bar DD^*$ molecule.  As we have already emphasized that this conclusion is not trivial, considering the near threshold $X(3872)$ resonance as a counterexample. The question of whether $Z_c$ is a bound state or a virtual one still remains open, though the latter is slightly preferred. Furthermore, we find that the main decay channels of $\zc$ are $J/\psi\pi$ (possibly including effects of other lighter channels) and  $D\bar D^*$. No strong evidence is found for its decay into $h_c\pi$.

It was recently emphasized in Refs.~\cite{Chen:2013coa,Liu:2013vfa} that, in a special kinematic region, obvious threshold enhancement was produced by a meson triangle diagram without introducing a genuine state for $Z_c(3900)$. Furthermore, it is suggested in Refs.~\cite{Liu:2015taa,Szczepaniak:2015eza} that the anomalous triangle singularity may also have a significant impact on the near-threshold behavior. We also investigate this interesting situation. Our  preliminary result shows that the ``anomalous threshold" may be of some importance to improve the fit quality but does not change the qualitative picture obtained in this paper, i.e., $Z_c(3900)$ is a molecule composed of $D\bar D^*$.

Finally we briefly comment on the recently discovered $P_c(4450)$ state~\cite{Aaij:2015tga}.
The $P_c(4450)$ state was observed by the LHCb Collaboration in the $J/\psi p$ channel in the $\Lambda_b^0\to J/\psi K^- p$ process. Since a proton is made of $uud$, the former process is very similar to the one under investigation if one replaces a $ud$ pair inside the proton by a $\bar d$ which are of same quantum number. Therefore one naturally expects the $P_c(4450)$ state to be a molecular state made of $\Sigma_c D^*$. Nevertheless, such a suggestion requires the spin quantum number of $P_c(4450)$ to be ${3\over 2}$ rather than ${5\over 2}$ as preferred by Ref.~\cite{Aaij:2015tga}. The investigation along this research line is ongoing.

\section*{Acknowledgments}
H.Q.Z would like to thank Yuan-Ning Gao for helpful discussions. This work is supported in part by National Nature Science Foundations of China (NSFC) under Contract Nos. 10925522, 11021092, 11575052 and 11105038, the Natural Science Foundation of Hebei Province with contract No.~A2015205205, the Sino-German Collaborative Research Center ``Symmetries and the Emergence of Structure in QCD'' (CRC~110) cofunded by the DFG and the NSFC.

\appendix{\center\bf\huge Appendix}

\section{ The loop integrals}\label{appendix.loop}
The $\bar{D}D^*$ meson loop integral  can be divided into transverse and longitudinal parts,
\begin{align}\label{eq11}
\Pi^{\mu\nu}=\int\frac{\mathrm{d^D}k}{\left(2\pi\right)^D}\frac{g^{\mu\nu}-\frac{k^\mu k^\nu}{m^2_{D^*}}}{\left(k^2-m^2_{D^*}\right)\left((p-k)^2-m^2_D\right)}=P^{\mu\nu}_T\Pi_T+P^{\mu\nu}_L\Pi_L,
\end{align}
and

\begin{align}\label{pit}
&\Pi_T=\frac{-i}{16\pi^2}\left(\frac{I_0}{2}-\frac{p^2}{2m^2_{D^*}}I_2+\frac{p^2+m^2_{D^*}-m^2_{D}}{2m^2_{D^*}}I_1-
\frac{\frac{1}{3}p^2-\left(m^2_{D^*}+m^2_D\right)}{4m^2_{D^*}}\right)\nonumber\\
&  -\frac{iR}{16\pi^2}\left(1+\frac{p^2}{12m^2_{D^*}}-\frac{m^2_{D^*}+m^2_D}{4m^2_{D^*}}\right),
\end{align}
\begin{align}\label{pil}
\Pi_L=\Pi_T+\frac{i}{16\pi^2}\frac{p^2}{m^2_{D^*}}I_2+\frac{iR}{16\pi^2}\frac{p^2}{3m^2_{D^*}},
\end{align}
where $R=-\frac{1}{\epsilon}+\gamma_E-\ln{4\pi}$ is the ultraviolet divergent part, and $p=p_{\bar{D}}+p_{D^*}$ .

The scalar integrals $I_n$ in Eqs.~(\ref{pit}) and (\ref{pil}) are
\begin{align}
  &I_n=\int_0^1\mathrm{d}x x^n\ln{\frac{m^2_Dx+m^2_{D^*}\left(1-x\right)-p^2x\left(1-x\right)}{\mu^2}}\,, \quad n=0,1,2.\nonumber\\
  &I_0=-B_0(s),\nonumber\\
  &I_1=-\frac{s+m_{D^*}^2-m_D^2}{2s}B_0(s)+\frac{A_0(m_{D^*})-A_0(m_D)}{2s},\nonumber\\
  &I_2=-\frac{\left(s+m_{D^*}^2-m_D^2\right)^2-m_{D^*}^2s}{3s^2}B_0(s)+\frac{A_0(m_{D^*})-2A_0(m_D)}{3s}\nonumber\\
  &-\frac{1}{18}+\frac{m_{D^*}^2-m_D^2}{3s^2}\left(A_0(m_{D^*})-A_0(m_D)\right)+\frac{m_{D^*}^2+m_D^2}{6s}.
\end{align}
Here the basic one-point loop function $A_0(m)$ and two-point loop function $B_0(s)$ are, respectively, 
\begin{align}
  &A_0(m)=-m^2\left(-1+\ln{\frac{m^2}{\mu^2}}\right)\,,\nonumber\\
  &B_0(s)=2-\ln{\frac{m_D^2}{\mu^2}}+\frac{s+m_{D^*}^2-m_D^2}{2s}
  \ln{\frac{m_D^2}{m_{D^*}^2}}+\rho_{m_D m_{D^*}}(s)\ln{\frac{\lambda(s)-1}{\lambda(s)+1}}\,,
\end{align}
where the kinematic factor $\lambda(s)=\sqrt{\frac{s-(m_{D^*}+m_D)^2}{s-(m_{D^*}-m_D)^2}}$ and
$\rho_{m_1m_2}(s)$ is given by
\begin{equation}\label{defrho}
\rho_{m_1m_2}(s)=\frac{\sqrt{\left(s-\left(m_1+m_{2}\right)^2\right)\left(s-\left(m_1-m_{2}\right)^2\right)}}{s}\,.
\end{equation}

Hence Eqs.~\eqref{pit} and \eqref{pil} could be rewritten as
\begin{align}
&-i\Pi_T(s)=\frac{1}{16\pi^2}(\frac{s-3\left(m_D^2+m_{D^*}^2\right)}{18m_{D^*}^2}-\frac{s+m_D^2-m_{D^*}^2}{12m_{D^*}^2s}A_0(m_D)
  - \frac{s+m_{D^*}^2-m_{D}^2}{12m_{D^*}^2s}A_0(m_{D^*})\nonumber\\
  &+\frac{s^2+m_{D^*}^4+m_D^4+10m_{D^*}^2s-2m_{D^*}^2m_D^2-2m_D^2s}{12m_{D^*}^2s}B_0(s)
  )-\frac{R}{16\pi^2}\left(1+\frac{s}{12m^2_{D^*}}-\frac{m^2_{D^*}+m^2_D}{4m^2_{D^*}}\right)\,,
\end{align}
\begin{align}
  &-i\Pi_L(s)=\frac{1}{16\pi^2}(\frac{\left(m_D^2-m_{D^*}^2-3s\right)A_0(m_D)}{4m_{D^*}^2s}+
  \frac{\left(-m_D^2+m_{D^*}^2+s\right)A_0(m_{D^*})}{4m_{D^*}^2s}\nonumber\\
  &-\frac{\left(m_D^4-2m_D^2\left(m_{D^*}^2+s\right)+\left(m_{D^*}^2-s\right)^2\right)B_0(s)}{4m_{D^*}^2s}
  )+\frac{R}{16\pi^2}\left(1+\frac{s}{4m^2_{D^*}}-\frac{m^2_{D^*}+m^2_D}{4m^2_{D^*}}\right)\,.
\end{align}

Several different approaches are adopted in the literature to renormalize the loops in the nonperturbative  calculation~\cite{Baru:2010ww,Artoisenet:2010va,Hanhart:2011jz,Guo:2016bjq}. In the present work, the divergences in Eqs.~\eqref{pit} and \eqref{pil} are removed through the substraction at the physical threshold:
\begin{eqnarray}
   \nonumber
     \bar{\Pi}_T(s) = \Pi_T(s)- \Pi_T(s_{th})\ ,\,\,\,     \bar{\Pi}_L(s) = \Pi_L(s)- \Pi_L(s_{th})\,,
\end{eqnarray}
with $s_{th}=(m_D+m_{D^*})^2$. Notice that choosing subtraction at a particular point (threshold) does not bring any additional constraint on the fits, since the effect of an arbitrary subtraction point can be absorbed by other fitted parameters. The near threshold behaviors of $\bar{\Pi}_T(s)$ and $\bar{\Pi}_L(s)$ are
\begin{eqnarray}
   \nonumber
     &\bar{\Pi}_T(s)=\frac{1}{16\pi}\left(-\rho_{m_D m_{D^*}}(s)+O(\rho^2(s))\right), \\
    &\bar{\Pi}_L(s)=\frac{1}{16\pi}\left(\rho_{m_D m_{D^*}}^3(s)+O(\rho^4(s))\right).
  \end{eqnarray}

Other relevant tensor integrals that appear in the expressions of decay amplitudes in the following discussions are given by
\begin{equation}
  \Pi_{1,\mu\nu\alpha}=\int\frac{\mathrm{d^D}k}{\left(2\pi\right)^D}\frac{\left(g_{\mu\nu}-\frac{k_\mu k_\nu}{m^2_{D^*}}\right)k_\alpha}{\left(k^2-m^2_{D^*}\right)\left(\left(p-k\right)^2-m^2_D\right)}\ ,
\end{equation}
\begin{equation}
  \Pi_{2,\mu\nu\alpha\beta}=\int\frac{\mathrm{d^D}k}{\left(2\pi\right)^D}\frac{\left(g_{\mu\nu}-\frac{k_\mu k_\nu}{m^2_{D^*}}\right)k_\alpha k_\beta}{\left(k^2-m^2_{D^*}\right)\left(\left(p-k\right)^2-m^2_D\right)}\ .
\end{equation}
After Feynman parametrization the above tensor integrals take the form
\begin{align}
  &\Pi_{1,\mu\nu\alpha}=\int_0^1\mathrm{d}x\left(\left(xp_\alpha g_{\mu\nu}-\frac{x^3p_\mu p_\nu p_\alpha}{m^2_{D^*}}\right)\Sigma_1-\frac{p_\nu x}{m^2_{D^*}}\Sigma_{2,\mu \alpha}-\frac{p_\mu x}{m^2_{D^*}}\Sigma_{2,\nu \alpha}\right)\ ,
\end{align}
\begin{align}
  &\Pi_{2,\mu\nu\alpha\beta}=\int_0^1\mathrm{d}x\left(p_\nu p_\beta\Sigma_{2,\mu \alpha}+
p_\nu p_\alpha \Sigma_{2,\mu \beta}
+p_\mu p_\beta \Sigma_{2,\nu\alpha}+p_\mu p_\alpha\Sigma_{2,\nu\beta}+p_\alpha p_\beta\Sigma_{2,\alpha\beta}+p_\mu p_\nu\Sigma_{2,\alpha\beta}\right)\nonumber\\
&\left(-\frac{x^2}{m^2_{D^*}}\right)+\int_0^1\mathrm{d}x\left(\left(g_{\mu\nu}p_\alpha p_\beta x^2-\frac{x^4p_\mu p_\nu p_\alpha p_\beta}{m^2_{D^*}}\right)\Sigma_1+g_{\mu\nu}\Sigma_{2,\alpha\beta}-\frac{1}{m^2_{D^*}}\Sigma_{3,\mu\nu\alpha\beta}\right)\ ,
\end{align}
where
\begin{align}\label{sigma1}
  \Sigma_1=\int\frac{\mathrm{d^D}k}{\left(2\pi\right)^D}\frac{1}{\left(k^2-\Delta\right)^2}=\frac{-i}{16\pi^2}\left(R+\ln{\Delta}\right)\ ,
  \end{align}
  \begin{align}\label{sigma2}
   \Sigma_{2,\mu\nu}=\int\frac{\mathrm{d^D}k}{\left(2\pi\right)^D}\frac{k_\mu k_\nu}{\left(k^2-\Delta\right)^2}=\frac{i}{16\pi^2}\frac{g_{\mu\nu}}{2}\Delta\left(\left(-R+1\right)-\ln{\Delta}\right)\ ,
   \end{align}
   \begin{align}\label{sigma3}
  &\Sigma_{3,\mu\nu\alpha\beta}=\int\frac{\mathrm{d^D}k}{\left(2\pi\right)^D}\frac{k_\mu k_\nu k_\alpha k_\beta}{\left(k^2-\Delta\right)^2}=\frac{i}{16\pi^2}\frac{\Delta^2}{8}\left(\left(-R+\frac{3}{2}\right)-\ln{\Delta}\right)\left(g_{\mu\nu}g_{\alpha\beta}+
  g_{\mu\alpha}g_{\nu\beta}+g_{\mu\beta}g_{\nu\alpha}\right),
\end{align}
and $\Delta=m^2_Dx+m^2_{D^*}\left(1-x\right)-p^2x\left(1-x\right)$. The divergences in Eqs.~(\ref{sigma1}), (\ref{sigma2}), (\ref{sigma3}) are also removed by the threshold subtraction.

\section{Decay amplitudes}\label{appendix.amp}
\subsection{ The amplitude of $X(4260)\rightarrow J/\psi\pi^{+}\pi^{-}$}
Based on the effective Lagrangians given in Sec.~\ref{sec.theo}, one can calculate the diagrams depicted in Fig.~\ref{zc3}. When summing the bubble chain diagrams, it is useful to distinguish whether the composite vertices in Fig.~\ref{zc3} depend on the loop momentum or not. For the composite vertices involving the external states, we use subscripts A and B to represent the loop momentum independent and dependent parts, respectively. The amplitude of $X(4260)\rightarrow J/\psi\pi^{+}\pi^{-}$  can be divided into four parts:
\begin{align}\label{eqjpsiaa}
 & iM_{1AA}^{\mu\nu}=\frac{\left(f_{1}-\frac{f_{5}g_{4}}{m_{Z}^{2}}\right)\left(\lambda_{2}-\frac{f_{5}g_{5}}{m_{Z}^{2}}\right)q\cdot q^{+}q^{-}\cdot q_{0}\Pi_{L}P_{L}^{\mu\nu}+f_{3}\lambda_{4}q^{+\mu}q^{-\nu}q_{\mu_{1}}q_{0\nu_{1}}\Pi_{L}P_{L}^{\mu_{1}\nu_{1}}}{1-i\left(\lambda_{1}-\frac{f_{5}^{2}}{m_{Z}^{2}}\right)\Pi_{L}}\nonumber\\
 &+
 \frac{\left(f_{1}+\frac{f_{5}g_{4}}{l^{2}-m_{Z}^{2}}\right)\left(\lambda_{2}+\frac{f_{5}g_{5}}{l^{2}-m_{Z}^{2}}\right)q\cdot q^{+}q^{-}\cdot q_{0}\Pi_{T}P_{T}^{\mu\nu}+f_{3}q^{+\mu}q_{\mu_{1}}\left(\lambda_{2}+\frac{f_{5}g_{5}}{l^{2}-m_{Z}^{2}}\right)q^{-}\cdot q_{0}\Pi_{T}P_{T}^{\mu_{1}\nu}}{1-i\left(\lambda_{1}+\frac{f_{5}^{2}}{l^{2}-m_{Z}^{2}}\right)\Pi_{T}}\nonumber \\
 &+\frac{\lambda_{4}q^{-\nu}q_{0\nu_{1}}\left(f_{1}+\frac{f_{5}g_{4}}{l^{2}-m_{Z}^{2}}\right)q^{+}\cdot q\Pi_{T}P_{T}^{\mu\nu_{1}}+f_{3}\lambda_{4}q^{+\mu}q^{-\nu}q_{\mu_{1}}q_{0\nu_{1}}\Pi_{T}P_{T}^{\mu_{1}\nu_{1}}}{1-i\left(\lambda_{1}+\frac{f_{5}^{2}}{l^{2}-m_{Z}^{2}}\right)\Pi_{T}}\nonumber\\
 &+\frac{f_{3}q^{+\mu}q_{\mu_{1}}\left(\lambda_{2}-\frac{f_{5}g_{5}}{m_{Z}^{2}}\right)q^{-}\cdot q_{0}\Pi_{L}P_{L}^{\mu_{1}\nu}+\lambda_{4}q^{-\nu}q_{0\nu_{1}}\left(f_{1}-\frac{f_{5}g_{4}}{m_{Z}^{2}}\right)q^{+}\cdot q\Pi_{L}P_{L}^{\mu\nu_{1}}}{1-i\left(\lambda_{1}-\frac{f_{5}^{2}}{m_{Z}^{2}}\right)\Pi_{L}}\ ,
  \end{align}
  \begin{align}
 &iM_{1AB}^{\mu\nu}=\left(\lambda_{3}g^{\nu\nu_{1}}q^{-\alpha}\Pi_{1,\nu_{1}\nu_{2}\alpha}-\lambda_{5}q^{-\nu_{1}}\Pi_{1,\nu_{2}\nu_{1}\ }^{\ \ \nu}\right)\nonumber \\
 & \left(\frac{\left(f_{1}+\frac{f_{5}g_{4}}{l^{2}-m_{Z}^{2}}\right)q\cdot q^{+}P_{T}^{\mu\nu_{2}}+f_{3}q^{+\mu}q_{\mu_{1}}P_{T}^{\mu_{1}\nu_{2}}}{1-i\left(\lambda_{1}+\frac{f_{5}^{2}}{l^{2}-m_{Z}^{2}}\right)\Pi_{T}}
 +\frac{\left(f_{1}-\frac{f_{5}g_{4}}{m_{Z}^{2}}\right)q\cdot q^{+}P_{L}^{\mu\nu_{2}}+f_{3}q^{+\mu}q_{\mu_{1}}P_{L}^{\mu_{1}\nu_{2}}}{1-i\left(\lambda_{1}-\frac{f_{5}^{2}}{m_{Z}^{2}}\right)\Pi_{L}}\right)\ ,
 \end{align}
 \begin{align}
 &iM_{1BA}^{\mu\nu}=-\left(f_{2}g^{\mu\mu_{1}}q^{+\alpha}\Pi_{1,\mu_{1}\mu_{2}\alpha}+f_{4}q^{-\mu_{1}}\Pi_{1,\mu_{1}\mu_{2}\ }^{\ \ \mu}\right)\frac{\left(\lambda_{2}-\frac{f_{5}g_{5}}{m_{Z}^{2}}\right)q_{0}\cdot q^{-}P_{L}^{\mu_{2}\nu}+i\lambda_{4}q^{-\nu}q_{0\nu_{1}}P_{L}^{\mu_{1}\nu_{2}}}{1-i\left(\lambda_{1}-\frac{f_{5}^{2}}{m_{Z}^{2}}\right)\Pi_{L}}
 \nonumber \\
 &-\left(f_{2}g^{\mu\mu_{1}}q^{+\alpha}\Pi_{1,\mu_{1}\mu_{2}\alpha}+f_{4}q^{-\mu_{1}}\Pi_{1,\mu_{1}\mu_{2}\ }^{\ \ \mu}\right)
 \frac{\left(\lambda_{2}+\frac{f_{5}g_{5}}{l^{2}-m_{Z}^{2}}\right)q_{0}\cdot q^{-}P_{T}^{\mu_{2}\nu}+i\lambda_{4}q^{-\nu}q_{0\nu_{1}}P_{T}^{\mu_{1}\nu_{2}}}{1-i\left(\lambda_{1}+\frac{f_{5}^{2}}{l^{2}-m_{Z}^{2}}\right)\Pi_{T}}\ ,
 \end{align}
 \begin{align}
 &iM_{1BB}^{\mu\nu}=f_{2}\lambda_{3}q^{+\alpha}q^{-\beta}g^{\mu\mu_{1}}g^{\nu\nu_{1}}\Pi_{2,\mu_{1}\nu_{1}\alpha\beta}+f_{2}\lambda_{5}q^{+}_{\alpha}q^{-\nu_{1}}
 g^{\mu\mu_{1}}\Pi_{2,\mu_{1}\nu_{1}\ \ }^{\ \ \alpha\nu}+
 f_{4}\lambda_{5}q^{+\mu_{1}}q^{-\nu_{1}}\Pi_{2,\mu_{1}\nu_{1}\ \ }^{\ \ \mu\nu}\nonumber \\
 & +q^{+\mu_{1}}q^{-\alpha}g^{\nu\nu_{1}}\Pi_{2,\mu_{1}\nu_{1}\ \alpha}^{\ \ \mu\ }f_4\lambda_{3}
 -i\left(\frac{\left(\lambda_{1}+\frac{f_{5}^{2}}{l^{2}-m_{Z}^{2}}\right)P_{T}^{\mu_{2}\nu_2}}{1-i\left(\lambda_{1}+\frac{f_{5}^{2}}{l^{2}-m_{Z}^{2}}\right)\Pi_{T}}+
 \frac{\left(\lambda_{1}-\frac{f_{5}^{2}}{m_{Z}^{2}}\right)P_{L}^{\mu_{2}\nu_2}}{1-i\left(\lambda_{1}-\frac{f_{5}^{2}}{m_{Z}^{2}}\right)\Pi_{L}}\right)\nonumber \\
 & \left(f_{2}g^{\mu\mu_{1}}q^{+\alpha}\Pi_{1,\mu_{1}\mu_{2}\alpha}+f_{4}q^{+\mu_{1}}\Pi_{1,\mu_{1}\mu_{2}\mu}\right)
 \left(\lambda_{3}g^{\nu_{1}\nu}q^{-\beta}\Pi_{1,\nu_{2}\nu_{1}\beta}-\lambda_{5}q^{-\nu_{1}}\Pi_{1,\nu_{2}\nu_{1}\ }^{\ \ \nu}\right)\ .
\end{align}
The momenta of $X(4260)$, $Z_c(3900)$, $J/\psi$ and $\pi^{+}(\pi^{-})$ are labeled as $q$, $l$, $q_0$ and $q^{+}(q^{-})$, respectively. $\epsilon_{X}^{\mu}$ and $\epsilon_{\psi}^{\mu}$ are the polarization vectors of $X(4260)$ and $J/\psi$, in order.
Finally, the full amplitude of $X(4260)\rightarrow J/\psi\pi^{+}\pi^{-}$ can be written as
\begin{align}\label{jpsi}
  iM_{X(4260)\rightarrow J/\psi\pi^{+}\pi^{-}}=\epsilon_{X\mu}\epsilon_{\psi\nu}^{*}\left(iM_{1AA}^{\mu\nu}+iM_{1AB}^{\mu\nu}+iM_{1BA}^{\mu\nu}+iM_{1BB}^{\mu\nu}\right).
\end{align}

In order to simplify the above and following expressions, we have redefined the coupling parameters to absorb the pion decay constant $f_\pi$ and its mass $m_\pi$.

\subsection{ The amplitude of $X(4260)\rightarrow \bar{D}D^{*}\pi^{\pm}$}
The amplitude of $X(4260)\rightarrow \bar{D}D^{*}\pi^{+}$ needs to be divided into two parts:
\begin{align}
iM_{2A}^{\mu\nu}=i\frac{\left(f_{1}+\frac{f_{5}g_{4}}{l^{2}-m_{Z}^{2}}\right)q\cdot q^{+}P_{T}^{\mu\nu}+f_{3}q^{+\mu}q_{\mu_{1}}P_{T}^{\mu_{1}\nu}}{1-i\left(\lambda_{1}+\frac{f_{5}^{2}}{l^{2}-m_{Z}^{2}}\right)\Pi_{T}}
+i\frac{\left(f_{1}-\frac{f_{5}g_{4}}{m_{Z}^{2}}\right)q\cdot q^{+}P_{L}^{\mu\nu}+f_{3}q^{+\mu}q_{\mu_{1}}P_{L}^{\mu_{1}\nu}}{1-i\left(\lambda_{1}-\frac{f_{5}^{2}}{m_{Z}^{2}}\right)\Pi_{L}},
\end{align}
\begin{align}
&iM_{2B}^{\mu\nu}=-if_{2}q_{D^{*}}\cdot q^{+}g^{\mu\nu}-if_{4}q_{D^{*}}^{\mu}q^{+\nu}
  +\left(f_{2}g^{\mu\mu_{1}}q^{+\rho}\Pi_{1,\mu_{1}\nu_{1}\rho}+f_{4}q^{+\mu_{1}}\Pi_{1,\mu_{1}\nu_{1}\mu}\right)\nonumber \\
  &\left(\frac{\left(\lambda_{1}+\frac{f_{5}^{2}}{l^{2}-m_{Z}^{2}}\right)P_{T}^{\nu_{1}\nu}}{1-i\left(\lambda_{1}+
  \frac{f_{5}^{2}}{l^{2}-m_{Z}^{2}}\right)\Pi_{T}}+\frac{\left(\lambda_{1}-\frac{f_{5}^{2}}{m_{Z}^{2}}\right)P_{L}^{\nu_{1}\nu}}{1-i\left(\lambda_{1}-\frac{f_{5}^{2}}{m_{Z}^{2}}\right)\Pi_{L}}\right)\ ,
\end{align}
where $q_{D^*}$, $l=q_{D^*}+q_{D}$ are the momenta of $D^*$ and $Z_c(3900)$, and $\epsilon_{X}^{\mu}$ and $\epsilon_{D^*}^{\mu}$ are the polarization vectors of $X(4260)$ and $D^*$, respectively. Then the full amplitude of $X(4260)\rightarrow \bar{D}D^{*}\pi^{+}$ can be written as
\begin{equation}\label{eqdd}
  iM_{X(4260)\rightarrow \bar{D}D^{*}\pi}=\epsilon_{X\mu}\epsilon_{D^{*}\nu}^{*}(iM_{2A}^{\mu\nu}+iM_{2B}^{\mu\nu})\ .
\end{equation}

\subsection{The amplitude of $X(4260)\rightarrow h_{c}\pi^{+}\pi^{-}$}
We divide the amplitude of $X(4260)\rightarrow h_{c}\pi\pi$ into four parts, which are given by
\begin{align}
 & iM_{3AA}^{\mu\nu}=\Pi_{L}\frac{\left(f_{1}-\frac{f_{5}g_{4}}{m_{Z}^{2}}\right)q\cdot q^{+}P_L^{\nu_{1}\mu}+f_{3}q^{+\mu}q^{\mu_{1}}P_L^{\mu_{1}\nu_{1}}}{1-i\left(\lambda_{1}-\frac{f_{5}^{2}}{m_{Z}^{2}}\right)\Pi_{L}}
\left(\lambda_{6}q_H^{\alpha}q_{\beta}^{-}\epsilon^{\nu\nu_{1}\alpha\beta}+if_{5}f_{7}q_H^{\rho}q_{\sigma}^{-}D_{\nu_1\beta}\epsilon^{\rho\sigma\beta\nu}\right)\nonumber\\
 &+\Pi_{T}\frac{\left(f_{1}+\frac{f_{5}g_{4}}{l^{2}-m_{Z}^{2}}\right)q\cdot q^{+}P_T^{\nu_{1}\mu}+f_{3}q^{+\mu}q^{\mu_{1}}P_T^{\mu_{1}\nu_{1}}}{1-i\left(\lambda_{1}+\frac{f_{5}^{2}}{l^{2}-m_{Z}^{2}}\right)\Pi_{T}}
 \left(\lambda_{6}q_H^{\alpha}q_{\beta}^{-}\epsilon^{\nu\nu_{1}\alpha\beta}+if_{5}f_{7}q_H^{\rho}q_{\sigma}^{-}D_{\nu_1\beta}\epsilon^{\rho\sigma\beta\nu}\right)\ ,
 \end{align}
 \begin{align}
 & iM_{3AB}^{\mu\nu}=\frac{\left(f_{1}+\frac{f_{5}g_{4}}{l^{2}-m_{Z}^{2}}\right)q\cdot q^{+}P_{T}^{\mu\mu_{1}}+f_{3}q^{+\mu}q^{\sigma}P_T^{\sigma\mu_{1}}}{1-i\left(\lambda_{1}+\frac{f_{5}^{2}}{l^{2}-m_{Z}^{2}}\right)\Pi_{T}}
 \left(-\lambda_{7}q_{\beta}^{-}\Pi_{1,\mu_{1}\nu_{1}\alpha}\epsilon^{\nu_{1}\nu\alpha\beta}\right)\nonumber \\
 &+\frac{\left(f_{1}-\frac{f_{5}g_{4}}{m_{Z}^{2}}\right)q\cdot q^{+}P_{L}^{\mu\mu_{1}}+f_{3}q^{+\mu}q^{\sigma}P_L^{\sigma\mu_{1}}}{1-i\left(\lambda_{1}-\frac{f_{5}^{2}}{m_{Z}^{2}}\right)\Pi_{L}}
 \left(-\lambda_{7}q_{\beta}^{-}\Pi_{1,\mu_{1}\nu_{1}\alpha}\epsilon^{\nu_{1}\nu\alpha\beta}\right)\ ,
\end{align}
\begin{align}
 &iM_{3BA}^{\mu\nu}=\left(f_{2}q^{+\alpha}\Pi_{1,\mu\nu_{1}\alpha}+f_{4}q^{+\mu_{1}}\Pi_{1,\mu_{1}\nu_{1}\ }^{\ \ \mu}\right)
 \left(-\lambda_{6}q_{H\alpha}q_{\beta}^{-}\epsilon^{\nu\nu_{1}\alpha\beta}-if_{5}\lambda_{6}q_{H\rho}q_{\sigma}^{-}D_{\alpha\beta}g^{\nu_{1}\alpha}\epsilon^{\rho\sigma\beta\nu}\right)
 \nonumber\\
 &+\left(f_{2}q^{+\alpha}\Pi_{1,\mu\mu_{2}\alpha}+f_{4}q^{+\mu_{1}}\Pi_{1,\mu_{1}\mu_{2}\ }^{\ \ \mu}\right)
 \left(-\lambda_{6}q_{H\alpha}q_{\beta}^{-}\epsilon^{\nu\nu_{1}\alpha\beta}-if_{5}\lambda_{6}q_{H\rho}q_{\sigma}^{-}D_{\alpha\beta}g^{\nu_{1}\alpha}\epsilon^{\rho\sigma\beta\nu}\right)\nonumber\\
 &\left(\frac{i\left(\lambda_{1}+\frac{f_{5}^{2}}{l^{2}-m_{Z}^{2}}\right)\Pi_{T}P_T^{\nu_{1}\mu_{2}}}{1-i\left(\lambda_{1}+\frac{f_{5}^{2}}{l^{2}-m_{Z}^{2}}\right)\Pi_{T}}+
 \frac{i\left(\lambda_{1}-\frac{f_{5}^{2}}{m_{Z}^{2}}\right)\Pi_{L}P_L^{\nu_{1}\mu_{2}}}{1-i\left(\lambda_{1}-\frac{f_{5}^{2}}{m_{Z}^{2}}\right)\Pi_{L}}\right)\ ,
\end{align}
\begin{align}
 &iM_{3BB}^{\mu\nu}=\left(\frac{i\left(\lambda_{1}+\frac{f_{5}^{2}}{l^{2}-m_{Z}^{2}}\right)P_{T}^{\mu_{2}\nu_{2}}}{1-i\left(\lambda_{1}+
 \frac{f_{5}^{2}}{l^{2}-m_{Z}^{2}}\right)\Pi_{T}}+\frac{i\left(\lambda_{1}-\frac{f_{5}^{2}}{m_{Z}^{2}}\right)P_{L}^{\mu_{2}\nu_{2}}}
 {1-i\left(\lambda_{1}-\frac{f_{5}^{2}}{m_{Z}^{2}}\right)\Pi_{L}}\right)\lambda_{7}q_{\beta}^{-}
 \Pi_{1,\nu_{2}\nu_{1}\alpha}\epsilon^{\nu_{1}\nu\alpha\beta}
 \nonumber\\
 &\left(f_{2}q^{+\alpha}\Pi_{1,\mu\mu_{2}\alpha}+f_{4}q^{+\mu_{1}}\Pi_{1,\mu_{1}\mu_{2}\ }^{\ \ \mu}\right)
+\left(f_{2}q^{+\sigma}\Pi_{2,\mu\nu_{1}\sigma\alpha}+f_{4}q^{+\mu_{1}}\Pi_{2,\mu_{1}\nu_{1}\ \alpha}^{\ \ \mu\ }\right)\lambda_{7}
 q^{-\beta}\epsilon^{\nu_{1}\nu\alpha\beta}\ ,
\end{align}
where $l$, $p_H$, $\epsilon_{X}^{\mu}$ and $\epsilon_{H}^{\mu}$ are the momenta of $Z_c(3900)$, $h_c$ and the polarization vectors of $X(4260)$, $h_c$, respectively. Then the full amplitude of $X(4260)\rightarrow h_{c}\pi\pi$ follows,
\begin{equation}\label{hc}
  iM_{X\rightarrow h_{c}\pi\pi}=\epsilon_{X\mu}\epsilon_{H\nu}^{*}(iM_{3AA}^{\mu\nu}+iM_{3AB}^{\mu\nu}+iM_{3BA}^{\mu\nu}+iM_{3BB}^{\mu\nu})\ .
\end{equation}

In order to implement the strong $\pi\pi$ final state interaction, we first need to project out the $s$-wave component of the  $\pi\pi$ system for the amplitudes of $J/\psi\pi\pi$ and $h_c\pi\pi$, and this is done by using  the helicity amplitude decomposition method in Ref.~\cite{Chung:1993da}. Then the final forms of the amplitudes after the implementation of the $\pi\pi$ final state interactions are given by Eq.~\eqref{fsipipi1}, see Ref.~\cite{Dai:2012pb} for details.

\section{The analysis of spectral density function sum rule}\label{sec.sdf}

\begin{figure}[htbp]
\centering
\includegraphics[width=0.6\textwidth]{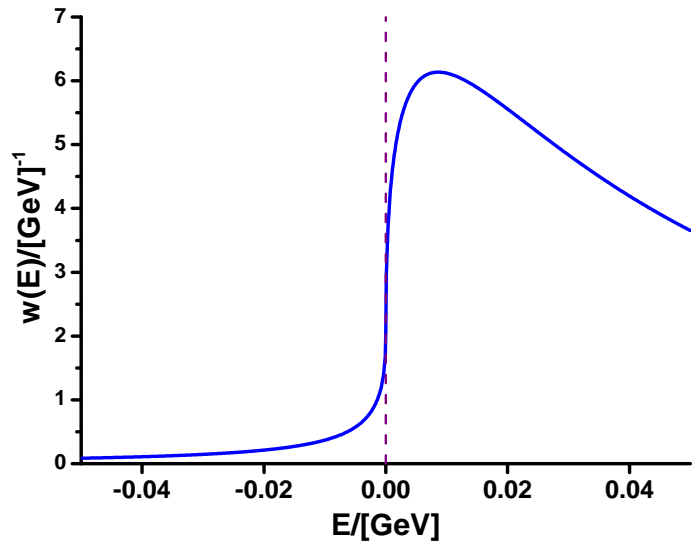}
\caption{The spectral density function of $Z_c(3900)$. }
\label{fig:SDF}
\end{figure}

We follow Refs.~\cite{Baru:2003qq,kala:2009na} to use the spectral density function method to calculate the probability for $Z_c(3900)$ to be an elementary state in this section. Both the $J/\Psi \pi$ and $D \bar D^*$ channels couple to the $Z_c(3900)$ state. We set $E=0$ at the $D \bar D^*$ threshold. For simplicity we omit the isospin violation and use the non-relativistic Flatt\'{e} parametrization of the SDF, which is analogous to the one adopted in Ref.~\cite{kala:2009na}
\begin{equation}\label{spdenmulc} 
w(E)=\frac{1}{2\pi}\frac{g_1 \sqrt{2\mu E}\theta(E)+\Gamma_0}{|E-E_0+\frac{i}{2}g_1 \sqrt{2\mu E}+\frac{i}{2}\Gamma_0|^2}\,,
\end{equation}
with $\mu=m_{D^{*-}}m_{D^0}/(m_{D^{*-}}+m_{D^0})$ the reduced mass.
We take from Fit II $M_{Z_c}=3.903$GeV and  $E_0=M_{Z_c}-(m_D+m_{\bar D^*})\sim 0.028$~GeV. Meanwhile, the decay width to the  inelastic channel is $\Gamma_0=0.01333$~GeV, cf. Table.\ref{table3}. The coupling constant $f_5$ in the Lagrangian ($\mathcal{L}\supset f_5 Z_c^{+\mu} D^{*-}_\mu D^0$) is $13.71$~GeV from Fit II. The coupling $g_1$ in Eq.~(\ref{spdenmulc}) can be obtained from $f_5$ as follows. The tree level decay width in the CM frame is
\[\Gamma(E)\sim \frac{1}{3}\sum_{\text{spins of }Z_c, D^*}|i f_5\epsilon^*_{D^*}\cdot \epsilon_{Z_c}|^2\frac{\sqrt{2\mu E}}{8\pi M_{Z_c}^2}
\sim \frac{f_5^2 \sqrt{2\mu E}}{8\pi M_{Z_c}^2}+O(E/M_{Z_c}^4)\,.\]
On the other hand, in the Flatt\'{e} parametrization the width is
\[\Gamma(E)\sim g_1 \sqrt{2\mu E}\,.\]
Therefore we obtain
\[g_1=\frac{f_5^2}{8\pi M_{Z_c}^2}\sim 0.491\,.\]
With the above inputs, we can then calculate the SDF, which is shown in Fig. \ref{fig:SDF}.

To proceed we calculate the possibility for $Z_c$ to be an elementary particle by taking the integral of the SDF in a reasonable energy interval~\cite{Baru:2003qq,kala:2009na}. In Fit II the central energy of $Z_c$ is about $E_c=4.000\times 10^{-3}$ GeV and the total width (pole width) is  $\Gamma=0.02984$ GeV. Then we integrate the SDF with different intervals and the results are given in Table \ref{tab:p}.

\begin{table}[htbp]
 \caption{Possibility ($\mathcal{Z}$) for $Z_c$ to be an elementary particle estimated with different integral intervals. }
 \label{tab:p}
 \begin{tabular}{ccccc}
  \hline\hline
   Integral interval & $[E_c-\Gamma/2,E_c+\Gamma/2]$ & $[E_c-\Gamma,E_c+\Gamma]$ & $[E_c-2\Gamma,E_c+2\Gamma]$ & $[E_c-3\Gamma,E_c+3\Gamma]$ \\
  \hline
 $\mathcal{Z}$ & 12.58\% & 20.62\% & 31.97\% & 39.67\% \\
  \hline\hline
 \end{tabular}
\end{table}

From Table~\ref{tab:p} we see that there are some ambiguities and uncertainties when estimating the elementariness coefficient  $\mathcal{ Z}$, since it is rather sensitive to the integral interval. Nevertheless, even if we take the interval as large as 6$\Gamma_c$, the $\mathcal{ Z}$ is still  smaller than $50\%$.\footnote{The contribution from the high energy tale of SDF might be overestimated here, since we do not consider the effect of form factor. Inclusion of the latter will reduce the uncertainty caused by the variation of the integral interval and decrease the magnitude of the $\mathcal{ Z}$ parameter, especially when the integration interval is large.} So we conclude that the SDF sum rule method confirms the molecular nature of $Z_c(3900)$.


\begin{thebibliography}{99}

\bibitem{Choi:2003ue} S.~K.~Choi {\it et al.} [Belle Collaboration], Phys.\ Rev.\ Lett.\  {\bf 91}, 262001 (2003).

\bibitem{Agashe:2014kda} K.~A.~Olive {\it et al.} [Particle Data Group Collaboration], Chin.\ Phys.\ C {\bf 38}, 090001 (2014).


\bibitem{Olsen:2014qna} C.~Z.~Yuan, Front. Phys. {\bf 10}, 101401 (2015).

\bibitem{Chen:2016qju} H.~X.~Chen, W.~Chen, X.~Liu and S.~L.~Zhu, Phys.\ Rept.\  {\bf 639}, 1 (2016). 


\bibitem{Weinberg:1962hj}S.~Weinberg, Phys.\ Rev.\  {\bf 130}, 776 (1963).

\bibitem{Baru:2003qq} V.~Baru et al., Phys.\ Lett.\ B {\bf 586}, 53  (2004).

\bibitem{Hyodo}T.~Hyodo, D.~Jido and A.~Hosaka, Phys.\ Rev.\ C {\bf 85}, 015201 (2012).


\bibitem{Aceti} F.~Aceti and E.~Oset, Phys.\ Rev.\ D {\bf 86}, 014012 (2012).


\bibitem{Agadjanov:2014ana}D.~Agadjanov et al., JHEP {\bf 1501}, 118 (2015).

\bibitem{Guo:2015daa}Z.~H.~Guo and J.~A.~Oller, Phys.~Rev.~{\bf D93}, 096001 (2016).

\bibitem{Chen:2015ata} W.~Chen et al., Phys.\ Rev.\ D {\bf 92}, 054002 (2015).

\bibitem{Zhang:2013aoa}J.~R.~Zhang, Phys.\ Rev.\ D {\bf 87}, 116004 (2013).

\bibitem{Wang:2013vex} Z.~G.~Wang and T.~Huang, Phys.\ Rev.\ D {\bf 89}, 054019 (2014).

\bibitem{Cui:2013yva} C.~Y.~Cui et al., J.\ Phys.\ G {\bf 41}, 075003 (2014).


\bibitem{Guo:2011pa}Z.~H.~Guo and J.~A.~Oller, Phys.\ Rev.\ D {\bf 84}, 034005 (2011).

\bibitem{Guo:2015dha} Z.~H.~Guo, U.~G.~Mei\ss ner and D.~L.~Yao, Phys.\ Rev.\ D {\bf 92}, 094008 (2015).

\bibitem{Lu:2016gev} J.~X.~Lu et al., Phys.\ Rev.\ D {\bf 93}, no. 11, 114028 (2016).

\bibitem{Morgan:1992ge}D. Morgan, Nucl. Phys. A {\bf 543}, 632 (1992); \\
K.~L~Au, D. Morgan and M. R. Pennington, Phys. Rev. D {\bf 35}, 1633 (1987).


\bibitem{Zhang:2009bv}O. Zhang, C. Meng, and H. Q. Zheng, Phys. Lett. B {\bf 680}, 453 (2009).

\bibitem{Meng:2014ota}C. Meng et al, Phys. Rev. D {\bf 92}, 034020 (2015).

\bibitem{Dai:2012pb}L. Y. Dai et al, Phys. Rev. D {\bf 92}, 014020 (2015).


\bibitem{Ablikim:2013mio}M.~Ablikim et al. (BESIII Collaboration), Phys. Rev. Lett. {\bf 110}, 252001 (2013).

\bibitem{Liu:2013dau}Z.~Q.~Liu et al. (Belle Collaboration), Phys. Rev. Lett. {\bf 110}, 252002 (2013).

\bibitem{Xiao:2013iha}T.~Xiao et al. (CLEO Collaboration),  Phys.\ Lett.\ B {\bf 727}, 366 (2013).


\bibitem{Ablikim:2013xfr}M. Ablikim et al. (BESIII Collaboration), Phys. Rev. Lett. {\bf 112}, 022001 (2014).



\bibitem{Maiani2013}L.~Maiani et al., Phys.\ Rev.\ D {\bf 87}, 111102 (2013).

\bibitem{Dias:2013xfa} J. Dias et al, Phys. Rev. D {\bf 88}, 016004 (2013).

\bibitem{Terasaki:2013lta}K. Terasaki, arXiv:1304.7080 [hep-ph].

\bibitem{Qiao:2013raa} C. F. Qiao and L. Tang, Eur. Phys. J. C {\bf 74}, 3122 (2014).


\bibitem{Braaten:2013boa} E. Braaten, Phys. Rev. Lett. {\bf 111}, 162003 (2013).

\bibitem{Navarra:2015rha} F. S. Navarra et al, Nucl. Part. Phys. Proc. {\bf 258}, 144 (2015).


\bibitem{Dong:2013iqa}Y. Dong et al, Phys. Rev. D {\bf 88}, 014030 (2013).

\bibitem{Guo:2013sya}F. K. Guo et al, Phys. Rev. D {\bf 88}, 054007 (2013).


\bibitem{He:2015mja}J. He, Phys. Rev. D {\bf 92}, 034004 (2015).

\bibitem{Wang:2013cya}Q. Wang, C. Hanhart, and Q. Zhao, Phys. Rev. Lett. {\bf 111}, 132003 (2013).

\bibitem{Wilbring:2013cha}E. Wilbring, H. W. Hammer, and U. G. Meissner, Phys. Lett. B {\bf 726}, 326 (2013).



 \bibitem{Guo:2013ufa}F. K. Guo, U. G. Meissner, and W. Wang, Commun. Theor. Phys. {\bf 61}, 354 (2014).


\bibitem{Zhou:2015jta}Z. Y. Zhou and Z. G. Xiao, Phys. Rev. D {\bf 92}, 094024 (2015).

\bibitem{Albaladejo:2015lob}  M.~Albaladejo et al., Phys.\ Lett.\ B {\bf 755}, 337 (2016).

\bibitem{Patel:2014vua}S.~Patel, M.~Shah and P.~C.~Vinodkumar, Eur.\ Phys.\ J.\ A {\bf 50}, 131 (2014).


\bibitem{Swanson:2014tra}E. S. Swanson, Phys. Rev. D {\bf 91}, 034009 (2015).

\bibitem{Chen:2013coa} D. Y. Chen, X. Liu, and T. Matsuki, Phys. Rev. D {\bf 88}, 036008 (2013).

\bibitem{Liu:2013vfa}X. H. Liu and G. Li, Phys. Rev. D {\bf 88}, 014013 (2013).

\bibitem{Ikeda:2016zwx} Y. Ikeda et al., Phys.\ Rev.\ Lett.\  {\bf 117}, no. 24, 242001 (2016).

\bibitem{Liu:2015taa}X. H. Liu, M. Oka, and Q. Zhao, Phys. Lett. B {\bf 753}, 297 (2016).


\bibitem{Szczepaniak:2015eza}A. P. Szczepaniak, Phys. Lett. B {\bf 747}, 410 (2015).


\bibitem{Chen:2014afa}Y. Chen et al, Phys. Rev. D {\bf 89}, 094506 (2014).

\bibitem{Li:2013xia}G. Li, Eur. Phys. J {\bf 73}, 2621 (2013).

\bibitem{Voloshin:2013dpa}M. B. Voloshin, Phys. Rev. D {\bf 87}, 091501 (2013).

\bibitem{Ke:2013gia}H. W. Ke, Z. T. Wei, and X. Q. Li, Eur. Phys. J. C {\bf 73}, 2561 (2013).


\bibitem{Lin:2013mka}Q. Y. Lin, X. Liu, and H. S. Xu, Phys. Rev. D {\bf 8}, 114009 (2013).


\bibitem{Guo:2014iya}F. K. Guo et al, Phys. Rev. D {\bf 91}, 051504 (2015).

\bibitem{Guo:2016bjq}
  F.-K.~Guo, C.~Hanhart, Y.~S.~Kalashnikova, P.~Matuschek, R.~V.~Mizuk, A.~V.~Nefediev, Q.~Wang and J.-L.~Wynen,
  Phys.\ Rev.\ D {\bf 93}, no. 7, 074031 (2016).

\bibitem{pavon.pwc.230216.1}M.~P.~Valderrama, Phys.\ Rev.\ D {\bf 85}, 114037 (2012).

\bibitem{Aceti:2014uea} F.~Aceti et al., Phys.\ Rev.\ D {\bf 90}, 016003 (2014).

\bibitem{mehen} E.~Braaten and M.~Kusunoki, Phys. Rev. D {\bf 69}, 074005 (2004).

\bibitem{Bijnens:1999sh}J. Bijnens, G. Colangelo, and G. Ecker, JHEP {\bf 9902}, 020 (1999).

\bibitem{Dai:2011bs}L.~Y.~Dai, X.~G.~Wang and H.~Q.~Zheng, Commun.\ Theor.\ Phys.\  {\bf 57}, 841 (2012).

\bibitem{Zhou:2004ms}Z.~Y.~Zhou et al., JHEP {\bf 0502}, 043 (2005).

\bibitem{Caprini:2005zr}I.~Caprini, G.~Colangelo and H.~Leutwyler, Phys.\ Rev.\ Lett.\  {\bf 96}, 132001 (2006).

\bibitem{Ablikim:2013wzq}M. Ablikim et al. (BESIII Collaboration), Phys. Rev. Lett. {\bf 111}, 242001 (2013).

\bibitem{Hanhart:2015cua}
  C.~Hanhart, Y.~S.~Kalashnikova, P.~Matuschek, R.~V.~Mizuk, A.~V.~Nefediev and Q.~Wang,
  Phys.\ Rev.\ Lett.\  {\bf 115}, no. 20, 202001 (2015).

\bibitem{Lees:2012cn}J. Lees et al. (BaBar Collaboration), Phys. Rev. D {\bf 86}, 051102 (2012).

\bibitem{kala:2009na}Yu. ~Kalashnikova and A.~V.~Nefediev, Phys. Rev. D {\bf 80}, 074004 (2009).




\bibitem{Suzuki:2005ha}M. Suzuki, Phys. Rev. D {\bf 72}, 114013 (2005).

\bibitem{Meng:2005er}C. Meng, Y. J. Gao, and K. T. Chao,  Phys. Rev. D {\bf 87}, 074035 (2013).

\bibitem{kala}Yu.~Kalashnikova and A.~V.~Nefediev, Phys. Atom. Nucl. {\bf 76}, 1533 (2013).

\bibitem{Aaij:2015tga}R. Aaij et al.(LHCB Collaboration), Phys. Rev. Lett. {\bf 115}, 072001 (2015).

\bibitem{Baru:2010ww}
  V.~Baru, C.~Hanhart, Y.~S.~Kalashnikova, A.~E.~Kudryavtsev and A.~V.~Nefediev,
  Eur.\ Phys.\ J.\ A {\bf 44}, 93 (2010).

\bibitem{Artoisenet:2010va}
  P.~Artoisenet, E.~Braaten and D.~Kang,
  Phys.\ Rev.\ D {\bf 82}, 014013 (2010).

\bibitem{Hanhart:2011jz}
  C.~Hanhart, Y.~S.~Kalashnikova and A.~V.~Nefediev,
  Eur.\ Phys.\ J.\ A {\bf 47}, 101 (2011).

\bibitem{Chung:1993da}S. U. Chung, Phys. Rev. D {\bf 48}, 1225 (1993), [Erratum: Phys. Rev. D {\bf 56}, 4419 (1997)].

\end{thebibliography}
\end{document}